%% file: gtg.tex
\documentclass[a4paper,11pt,onecolumn]{article}
\usepackage{latexsym}
\usepackage{amssymb, amsmath}
\usepackage{graphicx}
\usepackage{dcolumn}
\usepackage{bm}

\def\be{\begin{equation}}
\def\ee{\end{equation}}
\def\beq{\begin{eqnarray}}
\def\eeq{\end{eqnarray}}
\def\bn{\begin{eqnarray*}}
\def\en{\end{eqnarray*}}

\def\P{\Phi}
\def\p{\phi}

\def\a{\alpha}
\def\b{\beta}
\def\s{\sigma}
\def\S{\Sigma}
\def\d{\delta}

\def\g{\gamma}

\def\G{\Gamma}

\def\k{\kappa}

\def\si{\psi}

\def\e{\epsilon}
\def\n{\nu}
\def\m{\mu}
\def\r{\rho}

\def\ta{\tau}

\def\l{\lambda}

\def\cL{{\cal{L}}}

\def\cC{{\cal{C}}}
\def\cD{{\cal{D}}}
\def\cF{{\cal{F}}}

\def\cR{{\cal{R}}}
\def\sl{\not\!}
\def\slD{\not\!D}

\def\draftversion{Y}                % Y for draft, N for final version
\def\note[#1]#2{\message{(#1)}\if\draftversion
{\noindent\em[#2]\/}\fi}
\begin{document}
\title{GAUGE THEORY OF GRAVITY AND SUPERGRAVITY}
\author{Romesh K. Kaul \\
The Institute of Mathematical Sciences,\\
Chennai 600 113, India.\\
kaul@imsc.res.in}

\maketitle

\begin{abstract} 
We present a formulation of gravity  in terms 
of a theory based on complex $SU(2)$ gauge fields with a general 
coordinate invariant action functional quadratic in the field strength. 
Self-duality or anti-self-duality  of the field strength emerges 
as a constraint from the equations
of motion of this theory. This in turn leads to Einstein gravity equations for
a dilaton and an axion conformally coupled to gravity for the self-dual
constraint. The analysis has also been extended to $N=1$ and $~ 2$
super Yang-Mills theory of complex $SU(2)$ gauge fields. 
This leads, besides other equations of motion, to 
self-duality/anti-self-duality of generalized
supercovariant field-strengths. The self-dual case is  then shown to 
yield as its solutions $N~=~1,~2$  supergravity equations respectively.
\end{abstract}

\input section1

\input section2

\input section3

\input section4

\input section5

\vskip1.5cm
\input acknowledgements

\input references\end{document}

%% file: section1
\section{\bf Introduction}

Quest for a gauge theory description of Einstein's General Theory of 
Relativity (GTR) has a long history. Pioneering
attempts made by Utiyama, Kibble and Mandelstam are now about five
decades old \cite{ukm}. There is also another more recent  and famous
formulation of Ashtekar where gravity is described in terms of 
a new variable, the complex $SU(2)$ Sen-Ashtekar gauge connection 
\cite{ash}. This gauge field is the self-dual part of the spin connection
$\omega_\m^{a b}$ where self-duality is with respect to the Lorentz 
indices $[a b]$.
The action used is complex. It is linear in the field strength much 
in the  same manner as the standard Einstein-Hilbert or 
Hilbert-Palatini action. The formulation is entirely chiral in that
it  deals with the  local Lorentz representations  involving only
the chiral part of $SL(2,~C)$ and not 
its conjugate. A Hamiltonian formulation is set up in terms 
of phase space consisting  of spatial part of the Lorentz self-dual
spin connection and its canonically conjugate  
density-weighted spatial triad. A related formulation is that of
Plebanski \cite{pleb} where  again we have  a first-order 
Palatini-type action for complex 
general relativity  described in terms of a  spinor-valued
two-form $\Sigma^{AB}$,  an $SL(2,~C)$ one-form $\omega_{A B}$ 
which is identified with the Lorentz-self-dual part of the 
spin connection and a totally symmetric Lagrange multiplier
field $\Psi_{ABCD}$ where the latin letters $A,~B,~C,~D$ 
 denote two-component spinor indices. 
Ashtekar canonical formulation may be viewed as the $(3+1)$ decomposition of
the first-order formalism of Plebanski \cite{riccardo}. 

There have  also been other attempts to set up  a  gauge theory 
description of gravity. For example, there is 
an $SL(3)$ and diffeomorphism invariant Euclidean space action,
still linear in field strength,  presented by 't Hooft \cite{thooft}.

As is well known, it has been a long standing challenge
to set up a quantum theory of gravity. It is generally believed that
perturbative quantum general relativity set up in terms of  quantized 
corrections to a background metric and as a perurbative expansion
in dimensionful Newton's constant is  not renormalizable. 
Choice of a background metric fixes the coordinate system and thus
breaks general covariance. It is possible that difficulties faced are due
to the tools and methods  used so far. It is possible that a consistent 
perturbative quantum description may be possible if it is set up
in terms of a  quantization based on some other, more suitable, 
set of fields with an appropriate action functional and as
a pertubation in terms of a dimesionless coupling instead of
dimensionful Newton's constant. For this we have to first develope 
a classical description of Einstein's general relativity in terms 
of these fields. The Newton's constant should emerge as a parameter
in the space of solutions of such a theory.  It is, therefore, worth 
while to explore various possible action principles involving only
dimensionless couplings which yield Einstein's 
equations of the classical gravity as solutions of their equations
of motion.

Gravity actions with quadratic curvature terms have been 
discussed for many decades now. For example, DeWitt in 1960 
 had hoped that such  terms may provide a cure to the 
divergence problem \cite{DeWitt}.  One of the earliest
studies of gravity theory with action made of only quadratic
curvature terms, $ R_{\m\n}^{~~~\a\b} ~R_{\a\b}^{~~~\m\n}$~,
was the parallel displacement gauge theory
of Yang {\cite{yang}. There are two types of variational
principles that can be adopted. In the Einstein-Hilbert 
variational picture (also known as second order formalism)
where spacetime is Riemannian, the
quadratic action is to be varied with respect to 
metric $ g_{\m\n}$.  This  yields a fourth order differential equation 
of motion for the graviton field $h^{\m\n}$ defined as 
${\sqrt {-g}}~g^{\m\n}~ =~ \eta^{\m\n} $ $+ ~\kappa h^{\m\n}$.
In fact a general higher derivative theory of gravity with
$R^2$, $R$ and cosmological constant terms can be shown to be 
renormalizable \cite{stelle}. But such theories
generically imply nonunitarity due to a negative residue 
spin two pole ({\it i.e.} a ghost) in the bare propagator of 
the gravition field.  On the other hand, in the
Einstein-Palatini variational picture (also called  first
order formalism), the connection (not Riemann-Chritoffel 
connection) and the metric are independently varied. This 
leads, in general, to space-times with torsion 
\cite{hehlrep}. Equation of motion obtained by varying 
the action,  quadratic in curvature  tensor, with respect to metric 
leads to a constraint expressing the gravitational stress-energy 
tensor to be zero.
This equation is solved by double self-dual or
anti-self-dual curvature tensor \cite{Rsquare}. These 
are only second order differential equations 
for the metric.  In the next section we shall recast this 
theory in terms of a complex $SU(2)$ gauge field theory. 
Unlike Ashtekar theory which is described in terms of an 
action linear in complex $SU(2)$ field strength ~$F_{\m\n}^i$~, 
here we shall deal with  an action functional
quadratic in this field strength. This is much
in line with gauge theories used to describe other fundamental
interactions of particle physics. Both the complex gauge
field and the metric are taken to be  independent variables
in the action as in the Einstein-Palatini variational
principle.  Solutions of the equations of motion
fall in to two classes: those with self-dual and
with anti-self-dual field strengths.  These constraints
can be solved to write the metric in terms of 
field strength of gauge fields. Thus geometric quantities
are related to the gauge fields. 
Finally we shall be led to standard Einstein  equations
of motion  for gravity conformally coupled to a dilaton and 
an axion as a solution to the self-dual constraint.
There is no dimensionful parameter in the definition of the
gauge theory. However, a dimensionful parameter, to be identified
with Newton's constant, will emerge as a modulus of the space
of solutions of the equations of motion of this theory.

In Section 3, we shall extend the discussion to $N=1$ 
complex $SU(2)$ super Yang-Mills theory. 
The equations of motion imply self-duality or anti-self-duality
of the supercovariantized field strength.  Supergravity theory 
emerges as a solution of  self-dual case. Same structure
gets carried over to the case of $N=2$ super Yang-Mills theory, where
self-duality of a generalized covariant field strength, obtained
as a solution to the equation of motion, leads to $N=2$ supergravity
equations. This we discuss in Section 4. 
Some concluding remarks will follow in Section 5.

%% file: section2
\section{Complex ~ SU(2)~  gauge ~ theory  as~  a  theory  of 
gravity}

Consider a complex $SU(2)$ gauge field $A_\m^i$ ($i = 1, 2, 3$)
and metric $g_{\m \n}$ as
independent variable fields in the action:

\be
S ~ =~ {\frac \ta 4}~ \int d^4x ~e~ g^{\m \a}~ g^{\n \b}~ F_{\m \n}^i~ 
F_{\a \b}^i \label{action}
\ee

\noindent where $e^2 ~=~ g~ = ~det ~g_{\m \nu}~ <~ 0$ ~  and 
complex field strength is:
\bn
F_{\m \nu}^i ~ =~ \partial_\m ~A_\n^i ~ - ~\partial _\nu ~A_\m^i ~ -~ 
\e^{ijk} ~A_\mu^j ~A_\n^k
\en

\noindent Here $\ta$ is a dimensionless complex coupling constant.
The action is complex; we may wish to make it real by adding to it 
a conjugate action given as a functional of  ${\bar A}_\m^i$ which is  
complex conjugate of the gauge field $ A_\m^i$:
\bn
{\bar S} ~ =~ {\frac {\bar \ta} 4}~ \int d^4x ~{\bar e} ~g^{\m \a} ~g^{\n \b}~ 
{\bar F}_{\m \n}^i ~{\bar F}_{\a \b}^i
\en
where $ {\bar F}_{\m \nu}^i ~ =~ \partial_\m ~{\bar A}_\n^i ~ - ~\partial _\nu 
~{\bar A}_\m^i ~ -~ \e^{ijk}~ {\bar A}_\mu^j ~{\bar A}_\n^k $ and $\bar e ~= ~-e$.
However, in the following we shall work with the complex action $S$.

The action $S$ is invariant under complex $SU(2)$ gauge transformations and
also under general coordinate transformations. It contains no kinetic energy
term for the metric $g_{\m\n}$. In contrast to Ashtekar theory \cite{ash}
where the action functional is linear in field strength, it is quadratic 
here.
\subsection{Equations of motion}

The equations of motion are obtained by varying the action  $S$ above
with respect to the independent fields. Variation with respect to 
the gauge field $A_\m^i$ yields the Yang-Mills equation of motion: 

\be
D^\m ~(e~F_{\m \n}^i) ~=~0 \label{ymeq}
\ee
\noindent where the gauge covariant derivative is: $D_\mu ~\P^i $ $ = 
\partial_\m ~\P^i - \e^{ijk}~ A_\m^j~ \P^k$.
Next,  variation of the action with respect to the
metric $g_{\m \n}$ gives the second equation of motion, which is in fact
a constraint equation:
\be
T_{\m \n} ~ \equiv ~ F_{~\m}^{i~~\a}~F_{\n \a}^i
~-~{\frac 1 4}~ g_{\m \n}~ F_{\a \b}^i ~F^{i \a \b} ~ =~ 0 \label{T}
\ee

\noindent Notice that gauge field stress-energy tensor $T_{\m \n}$ is traceless
$g^{\m \n}~ T_{\m \n} $ $ = 0$ and also  conserved ${\nabla }^\m ~T_{\m \n}
~=~ 0$ by the first of the two equations of motion, Yang-Mills equation 
(\ref{ymeq}).  Here  the derivative ${\nabla}_\m$
is covariant  with respect to the general coordinate
transformations: $ {\nabla}_\m 
~T^{\a \b} $ $= ~\partial_\m ~T^{\a \b} $$ +~ {\G}_{\m \l}^{~~~\a}
~T^{\l \b}  $ $ +~ {\G}_{\m \l}^{~~~\b}~T^{\a \l}$, where 
the Riemann-Christoffel
connection is given in terms of  metric through the condition:
$ \nabla_\m ~g_{\a \b} $$ \equiv \partial_\m ~g_{\a \b} $ 
$- \G_{\m \a}^{~~ ~\l} ~g_{\l \b} $
$- \G_{\m \b}^{~~~\l}~ g_{\a \l} $
$ =~0$.

We need to solve these equations of motion. To solve the constraint equation
(\ref{T}), introduce the dual field strength
\be
 ^*F^{i\mu \nu} ~ \equiv  ~{\frac {1} {2e}}
~\epsilon^{\mu \nu \alpha \beta} ~F^i_{\alpha \beta} 
\ee
\noindent where $\epsilon^{\mu \nu \alpha \beta}$ is the usual completely
antisymmetric Levi-Civita  density of weight one with values $\pm1$ or $0$. 
Notice  this duality operation is involutive: $^*(^*F_{\m\n}^i )$ $ =~ 
F_{\m\n}^i$ .
With this definition and the identity (in four dimensions)~
$ {\delta}_\mu ^{[\nu} \epsilon^{ \alpha \beta \rho \sigma]}  =  0$
(where square brackets indicate antisymmetrization of the contained
indices), it is straight forward to check that gauge field stress-energy
tensor can be rewritten as
\be
T_{\mu \nu}  ~\equiv ~ {\frac 1 2}~ ( F^i_{\mu \alpha} ~ + ~^*F^i_{\mu \alpha})
~(F^{i ~~\alpha}_{~\nu} ~ - ~ ^*F^{i~~\alpha}_{~\nu} ) 
\ee
\noindent Thus the constraint equation $T_{\m\n}~=~0$ is solved by 
self-dual or anti-self-dual field strength:~
\be
 F^i_{\mu \nu} ~ =~  \pm ~^*F^i_{\mu \nu} \label{SAD}
\ee
It is  important to notice that such  
field strengths satisfy the Yang-Mills equation of motion (\ref{ymeq})
identically. Also, for these field strengths, the
Lagrangian density becomes a total divergence:
\bn
 {\frac 1 4} ~e~ F^{i \mu\nu} ~ F^i_{\mu \nu} = \pm~{\frac 1 4}~ e~ ^*F^{i \mu
\nu} ~F^i_{\mu \nu}
 = \pm~{\frac 1 8}~  \epsilon^{\mu \nu \alpha \beta}
~F^i_{\mu \nu}~ F^i_{\alpha \beta} = \pm ~\partial_{\mu} ~J^{\mu}
\en
\noindent where  $ J^{\mu}~ =~ {\frac 1 2}~ \epsilon^{\mu \nu \alpha \beta}
~( A^i_\nu ~\partial_\alpha ~A^i_\beta ~- ~{\frac 1 3}~ \epsilon^{i j k} ~A^i_\nu
~A^j_\alpha ~A^k_\beta )$.

Self-duality or anti-self-duality constraint implies that the metric is
not an independent field, but can be solved for as a function of
the gauge field $A^i_\mu$ (or more exactly as a function of the gauge
field strength $ F^i_{\mu \nu}$). In fact, it can be shown that the metric
for (non-zero) self-dual  or anti-self dual field strength is
given by Urbantke type formulae \cite{U}: 
\be
 g^{-\frac 1 4 } ~ g_{\mu \nu} ~=~ ( det~ \phi_{i j})^{-\frac 1 2}
~X_{\mu \nu}, ~~~~~ 
g^{\frac 1 4} ~g^{\mu \nu} ~=~ ( det~ \phi_{i j})^{-\frac 1 2}
~Y^{\mu \nu} \label{U} 
\ee
\noindent where quantities $\p_{ij}$, $X_{\m\n}$ and $Y^{\m\n}$ are given in terms of
the self-dual  or anti-self-dual field strength as 
\bn
\phi_{i j} ~= ~\pm~{\frac 1 2} ~\epsilon^{\mu \nu \alpha \beta} ~F^i_{\mu
\nu}~ F^j_{\alpha \beta} 
\en
\noindent and 
\bn
X_{\mu \nu} &=& \pm~{\frac 2 3}~ \epsilon^{\alpha \beta \sigma \delta}
~\epsilon^{ i j k} ~ F^i_{\mu \alpha} ~F^j_{\beta \sigma}~ F^k_{\delta \nu} \cr 
 ~Y^{\mu \nu} &=& {\frac 1 3}~ \epsilon^{\mu \alpha \beta \gamma}
~\epsilon^{\lambda \nu \rho \sigma} ~\epsilon^{i j  k} ~F^i_{\beta \gamma}
~F^j_{\alpha \lambda}~ F^k_{\rho \sigma} 
\en

\noindent Thus (\ref{U}) gives the metric in terms of the 
self-dual or anti-self-dual field strength, but only  up to 
a conformal factor. This is so because self-duality
and anti-self-duality constraints are  not 
sensitive to the conformal factor of metric. Under a conformal
transformation ~$g_{\mu \nu} ~\rightarrow ~\Omega^{-2}~ g_{\mu \nu}$ ~ 
$(e ~\rightarrow~ \Omega^{-4}~e,$ $ ~ g^{\m\nu}~ \rightarrow~ \Omega^2~ g^{\m\n})$:
\bn
\left(F^{i \mu \nu} ~\mp~{\frac 1 {2e}} ~\epsilon^{\mu \nu \alpha \beta}
~F^i_{\alpha \beta} \right) \rightarrow
~\Omega^4 ~\left( F^{i \mu \nu}~ \mp ~{\frac 1 {2e}}~ 
\epsilon^{\mu \nu \alpha \beta}
~F^i_{\alpha \beta}\right)
\en
 
To analyse the self-duality or anti-self-duality constraint further, 
we trade the three
complex two-tensors $F_{\m\n}^i$ by six real two-tensors 
${\cR}_{\m\nu}^{~~~\a\b}$ for six values of the antisymmetric pair
of indices $(\a \b)$ through the definition:
\be
F^i_{\m \n} ~ = ~ {\cal R}_{\m\n}^{ ~~~\a \b}~ ~\Sigma^i_{\a \b} \label{defR}
\ee
where $\Sigma^i_{\mu \nu}$ is a self-dual two-form, $ ^* \Sigma^i_{\mu \nu}$
$ = \Sigma^i_{\mu \nu}$, which can be viewed as a curved space
generalization of the  flat Minkowski space   $\eta^i_{\m\n}$~-symbol  of 
't Hooft \cite{thooft1, thooft}. This is constructed from 
the tetrads $ e^a_\alpha$
defined as the square root of the metric through $ g_{\mu \nu} ~ =  
~e^a_\mu~ e^b_\nu ~\eta_{a b},$ ~~ $\eta_{a b} ~ = ~\delta_{a b} ~ =  ~
\eta^{a b} $ and $ a$  and $b$  take values  $ 1,2, 3, 4$. (In our notation, 
$ e^4_\alpha$ is pure imaginary). And 
$\Sigma^i_{\alpha \beta}  ~= $ $~\Sigma^{~~~i4}_{\alpha \beta} ~ +
~{\frac 1 2}~\epsilon^{ijk} ~\Sigma^{~~~jk}_{\alpha \beta}$ 
~ (each of $i,j,k$ takes values $1, 2,3$) 
where $\Sigma^{~~~ab}_{\alpha \beta}$ 
is the antisymmetrized product of tetrads $ e^a_\alpha$: 
$ \Sigma^{~~~ab}_{\alpha \beta}  \equiv  {\frac 1 2}~ 
e^a_{[\alpha }~e^b_{\beta]} $.

Self-duality or anti-self-duality of the field strength (\ref{SAD}) then implies the constraint:

\be
^*{\cR}_{\mu \nu}^{~~~a b} ~ = ~  \pm~\tilde{\cR}_{\mu \nu}^{~~~a b} \label{SD1} 
\ee
where $^*$ is the duality with respect to the first pair of indices $[\m \n]$
and tilde $\tilde{} $ is the duality in the second pair of flat  internal
space indices $[a b]$ defined as:  
\bn
 {\tilde{X}}^{ab}~  =~  {\frac 1 2} ~\epsilon^{a b c d}~ X_{c d}
\en
Here $\epsilon^{a b c d} $ is completely antisymmetric with 
$\epsilon^{ 1 2 3 4} = + 1$. The condtion (\ref{SD1})
is the double self-duality/anti-self-duality condition 
studied in references \cite{Rsquare}.

Next we use the  Lanczos identity:
\be
 ^*{\tilde{\cR}}_{\mu \nu}^{~~~a b} ~ \equiv ~ {\cR}^{a b}_{~~~\mu \nu}
~ +~  \Sigma^{~~~a b}_{\mu \nu}~ {\cal R} ~ + ~ 2 \Sigma_{\mu \nu}^{~~~c [a}
~{\cR}^{b]}_{~~~c} \label{identity}
\ee
where ${\cR}_\mu^{~~a}  ~= ~ {\cR}_{\mu \nu}^{~~~a b} ~e^\nu_b$
and ${\cal R}  =  {\cal R}_\mu^{~~a} ~e^\mu_a$.
This identity and self-duality or anti-self-duality of the field strength  
implies:
\be
\pm~{\cal R}_{\mu \nu}^{~~~a b} ~ -~ {\cal R}^{a b}_{~~~\mu \nu} ~ = ~
\Sigma^{~~~a b}_{\mu \nu} ~{\cR}~  - ~ e^{[a}_{[\mu} ~{\cR}^{b]}_{~~\nu]} 
\label{SD2}
\ee

We need to solve these constraints. To develop such  solutions we write:
\be
A^i_{\mu} ~= ~a^i_{\mu} ~+ ~b^i_{\mu} 
\ee
\noindent where $a^i_{\mu}$ is such that
$ D_{[\m} (a) ~\Sigma^i_{\n \a]}$ $ \equiv \partial_{[\mu} 
~\Sigma^i_{\nu \alpha]}~
$ $  - ~\epsilon^{ i j k} ~a^j_{[\mu} ~\Sigma^k_{\nu \alpha]} $ $~= ~0 $.
\noindent This constraint can be solved for $a^i_{\mu}$ in terms of
the tetrads through $\Sigma^i_{\mu \nu}$ to obtain:
\be
a^i_{\mu} ~= ~\omega_{\mu \alpha \beta} (e)~ \Sigma^{i \alpha \beta}\label{a+b}
\ee
where $\omega (e)$ is the usual spin-connection given in terms of the tetrads
$e^a_\m$:
\bn
\omega_{\mu \alpha \beta} (e) ~= ~{\frac 1 2} ~\left( e_{a \alpha}~
\partial_{[\beta }~e^a_{\mu]}~ + ~e_{a \beta}~ \partial_{[\mu}~
e^a_{\alpha]} ~- ~e_{a \mu}~ \partial_{[ \alpha}~e^a_{\beta ]}\right)
\en
Notice $a^i_\mu$ is the Sen-Ashtekar gauge field.

Next, we write 
\be
F^i_{\mu \nu} ~= ~f^i_{\mu \nu} ~+ ~\ell^i_{\mu \nu} \label{f+l}
\ee
where $f^i_{\m\n}$ is the field strength for the gauge field $a_\m^i$:
\bn
f^i_{\mu \nu} ~= ~\partial_{[\mu} ~a^i_{\nu]} ~- ~\epsilon^{ i j
k}~a^j_{\mu}~a^k_{\nu} ~\equiv ~R_{\mu \nu}^{~~~\alpha \beta}(\omega(e))~
\Sigma^i_{\alpha \beta}
\en
\noindent and 
\bn
 \ell^i_{\mu \nu} ~= ~D_{[\mu} (a) ~b^i_{\nu]} ~-~\epsilon^{ i j
k}~ b^j_{\mu}~b^k_{\mu} ~\equiv ~r_{\mu \nu}^{ ~~~\alpha \beta}~\Sigma^i_{\alpha
\beta}
\en
Here the derivative $D_\m (a)$ is gauge covariant derivative
involving the gauge field $a_\m^i$, $D_\m (a)~b^i_\n~=~ \partial_\m ~b^i_\n
~ -~ \epsilon^{ijk}~a_\m^j ~b_\n^k$;~ $R_{\mu \nu}^{~~~\alpha \beta}(\omega(e))$ 
is usual Riemann tensor and 
\be
\cR_{\m \n}^{~~~\a \b} ~=~ R_{\m \n}^{~~~\a \b}
(\omega(e)) ~ +~ 
r_{\m \n}^{~~~\a \b} \label{R+r}
\ee
where  tensor $r_{\mu \nu}^{~~~\alpha \beta}$ is 
to be determined.
Writing $ b^i_{\m} = h_{\m}^{~~\alpha \beta} ~\Sigma^i_{\alpha
\beta}$, this $r$-tensor is:
\be
r_{\mu \nu \alpha \beta} ~=~ \nabla_{[\mu}~h_{\nu] \alpha \beta}
~ + ~h_{\mu \alpha}^{~~~\lambda}~h_{\nu \lambda \beta}~ -~h_{\nu
\alpha}^{~~~\lambda}~h_{\mu \lambda \beta}\label{rtensor} 
\ee
Notice we have traded three complex vectors $A^i_{\mu}$
with six real vectors $h_{\m}^{~~\a \b}$ with six values of the
antisymmetric pair $(\a \b)$.   These can be viewed as contortion. 
The $24$ dimensional
space of real contortion $h_{\m\a\b}$ can be decomposed into three
irreducile subspaces: trace part  $h_\a ~=~ g^{\m\b}~h_{\m \a \b}$~,
completely antisymmetric part $K_{\m\a\b}$ and 
tensor part $J_{\m\a\b}$ with $g^{\m\b} J_{\m\a\b} =0 $ and
$J_{[\m\a\b]} = 0$ . These subspaces are respectively $4$, $4$ and
$16$ dimensional. In the following we shall take the tensor
part to be zero. Thus we parametrize  $h_{\mu \alpha \beta}$ as
\be 
h_{\mu \alpha \beta} ~=~ K_{\mu \alpha \beta}~ - ~{\frac 1 3}~
(g_{\mu \alpha}~h_{\beta}~ -~g_{\mu \beta}~h_{\alpha}) \label{h}
\ee
The four-tensor $r_{\m \n \a \b}$ is given by
\beq
r_{\m\n}^{~~~\a\b} = &&\nabla_{[\m} K_{\n]}^{~~\a\b} + ~{\frac 1 3}~ 
\d_{[\m}^{[\a} ~\nabla_{\n]} ~h^{\b]} ~+~ K_{[\m}^{~~\a\l}~K_{\n]\l}^{~~~\b}
-~{\frac 1 3} ~\d_{[\m}^{[\a} ~K_{\n] \l}^{~~~\b]} ~ h^\l \cr~ 
&&+~ {\frac 2 3} ~K_{\m\n}^{~~~[\a} ~h^{\b]} ~ +~ {\frac 1 9}
~\left( \d_{[\m}^{[\a} ~h_{\n]} ~h^{\b]} ~-~ \d_{[\m}^\a ~\d_{\n]}^\b ~h^2 \right)
\label{rfourtensor}~~~~~
\eeq
From this we  construct the two-tensor
$r_{\m \n} = r_{\m \a \n \b} ~ g^{\a \b}$.
The symmetric and antisymmetric parts of this tensor  and its trace 
($r~=~ g^{\m\n} r_{\m\n}$) are:
\beq
r_{\mu \nu} ~+ ~r_{\nu \mu} &= &{\frac 2 3} \left[\nabla_\mu ~h_\nu~ +
~\nabla_\nu ~h_\mu ~+ ~g_{\mu \nu} \nabla\cdot h
~ - ~{\frac 2 3}~(g_{\mu \nu} h^2 ~ -~ h_\mu h_\nu)\right] \cr
&&~-~2 ~K_{\mu \alpha \beta} ~K_\nu^{~~\alpha \beta} \cr
 r_{\mu \nu} ~- ~r_{\nu \mu} &= &{\frac 2 3} ~ \nabla_{[\mu} ~h_{\nu]}
~-~2 ~\nabla_\sigma ~K_{\mu \nu}^{~~~\sigma}\cr 
r &=& 2~\nabla \cdot h ~-~{\frac 2 3}~h^2 ~-~ K_{\m\a\b}~K^{\m\a\b} 
\label{rtensor1}
\eeq

Let us now consider the two cases of self-duality and anti-self-duality
separately.
\subsection{Self-dual solution}

\noindent Contracting constraint equation (\ref{SD2}) by $ e^\nu_b$ 
for the self-dual case yields:
\be
{\cR}_{\mu \nu} ~ +~  {\cR}_{\nu \mu} ~ = ~ {\frac 1 2}~ g_{\mu \nu}~
{\cR} \label{master}
\ee
This is our master equation which we wish to solve. It fixes nine of the 21
independent components of $\cR_{\m\n}^{~~~ab} + \cR^{a b}_{~~~\m\n}$
leaving 12 independent components undetermined. This equation when
substituted back in to (\ref{SD2}) yields the constraint
\be
\cR_{\mu \nu}^{~~~a b} ~ -~ \cR^{a b}_{~~~\mu \nu} ~ = ~
{\frac 1 2}~e_{[\m}^{[a}~ \left( \cR_{\n]}^{~~b]} ~ -~
\cR^{b]}_{~~\n]} \right) \label{SD3}
\ee
This equation fixes nine of the 15 independent components of
$\cR_{\m\n}^{~~~ab} - \cR^{a b}_{~~~\m\n}$ leaving 6 undetermined.
The two equations (\ref{master}) and (\ref{SD3}), which are equivalent
to the self-dual constraint (\ref{SD2}), then fix 18 of
the independent components of $\cR_{\m\n}^{~~~a b}$, other 18
are undetermined. Solving these would be
equivalent to solving the self-duality equation for our
$SU(2)$ gauge field strength (\ref{SAD}).

The constraint  (\ref{SD3})  further imlpies
\beq
\left( \nabla^{[\a} ~+ ~{\frac 2 3}~h^{[\a} \right)~ K^{\b]}_{~~\m\n} &-&  
\left( \nabla_{[\m} ~ + ~{\frac 2 3}~h_{[\m} \right)~ K_{\n]}^{~~\a\b} ~~~~~~\cr
&= &\left(\nabla_\s ~+ ~{\frac 2 3} ~h_\s \right) 
~\d_{[\m}^{[\a} ~K_{\n]}^{~~\b]\s} \label{Kconst}
\eeq
where we have used $\cR_{\m \n}^{~~~\a \b}$ $ = R_{\m \n}^{~~~\a \b}
(\omega(e)) $ $ + ~r_{\m \n}^{~~~\a \b}$ and the fact that Riemann 
tensor is symmetric under the interchange of  first and second 
pairs of indices 
and also therefore the Ricci tensor $R_{\m\n}(\omega (e))$ is symmetric. 
Next from (\ref{master}) we may write:
\bn
 R_{\mu \nu}(\omega (e)) ~= ~- ~ {\frac 1 2}~[r_{\mu \nu} ~+ ~r_{\nu \mu}] ~+ 
~{\frac 1 4}~g_{\mu \nu} ~[ R (\omega (e))~+ ~r] 
\en
Or equivalently
\beq
R_{\mu \nu}(\omega (e)) ~-~ {\frac 1 2}~ g_{\mu \nu} ~R (\omega (e))
=  -~t_{\mu \nu}  &\equiv& 
 - {\frac 1 2} ~\left[ r_{\mu \nu} ~+ ~r_{\nu \mu} ~- ~{\frac 1 2}~g_{\mu \nu} 
~ r \right] \cr
&& ~~~~~- ~{\frac 1 4} ~ g_{\mu \nu} ~ R(\omega (e)) \label{t}
\eeq
 Now since $\nabla^{\mu} ~[R_{\mu \nu}(\omega (e)) ~-
~{\frac 1 2}~ g_{\mu \nu} ~R(\omega (e))] \equiv 0$, we need to solve
$\nabla^\mu ~t_{\mu \nu}$ $ = 0$. This is what we shall attempt to do next.

\subsubsection{Dilaton-axion gravity from self-dual solution}

We shall make convenient ansatz for $h_\m$ and $K_{\m \a \b}$ in 
(\ref{h}):

\be
 h_\mu ~= ~-~3~ \partial_\mu \phi~, ~~~~~~  K_{\mu \alpha \beta}~ 
=  ~{\frac \kappa { 2\sqrt{2}}}~ e^{-2\phi}~ H_{\mu \alpha \beta} \label{ansatz}
\ee
where $\kappa$ is a ~constant and ~  completely ~antisymmetric ~$H_{\m \a \b}$ 
~ is the field ~  strength of an antisymmetric tensor gauge field $B_{\m\n}$:~
$ H_{\mu \alpha \beta} $ $ = \partial_{[\mu} ~B_{\alpha \beta]}$.
We shall take $\phi$ to be dimensionless (soon we shall see that it
will represent a dilaton) and antisymmetric gauge field $B_{\m\n}$
to have mass dimensions $+1$ and its field strength $H_{\m\a\b}$
then has dimensions
$+2$. In order mass dimension of $h_{\m\a\b}$ of (\ref{h}) be
$+1$ (mass dimension of the gauge field $A_\m^i$), the constant 
$\kappa$ has to be of dimension $-1$. However 
in the following discussion, we shall take  
$\kappa$ $ = 1$ for convenience; it can easily be restored whenever
needed.

We use (\ref{ansatz}) to construct the tensors $ r_{\m\n\a\b}$
and $r_{\m\n}$ of (\ref{rtensor}) and (\ref{rtensor1}). This leads
us to 
\beq
r_{\m\n}~=~&& 2~ \left[ \nabla_\m\p~\nabla_\m\p~ -~ \nabla_\m\nabla_\n \p
\right] ~-~ g_{\m\n}~ \left[2~(\nabla \p)^2 ~+~\nabla^2 \p \right] \cr
&& -~ {\frac 1 {2\sqrt{2}}}~~ \nabla_\a (e^{-2\p}~H_{\m\n}^{~~~\a})
~-~ {\frac 1 8}~e^{-4\p}~ H_{\m\a\b}~H_\n^{~~\a\b} \cr
r ~=~ && -6~\left[(\nabla \p)^2 ~+~ \nabla^2 \p \right] ~-~ 
{\frac 1 8} ~ e^{-4\p}~H_{\a\b\g}~H^{\a\b\g}
\eeq
From  (\ref{t}) these in turn imply
\beq
t_{\mu \nu}  ~&\equiv &-  
~2 ~\left[ \nabla_\mu \phi ~\nabla_\nu \phi ~- ~\nabla_\mu
\nabla_\nu \phi\right]\cr 
~&& -~{\frac 1 2}~ g_{\mu \nu}~ \left[ (\nabla \phi)^2
~-~\nabla^2 \phi  ~-~{\frac 1 2}~ R(\omega (e))\right] \cr
~&&-~{\frac 1 8}~ e^{-4\phi} ~\left[ H_{\mu \alpha \beta}
~H_\nu^{~~\alpha \beta} ~- ~{\frac 1 4}~ g_{\mu \nu} ~
H_{\alpha \beta \gamma} ~H^{\alpha \beta \gamma}\right] \label{t1}
\eeq
Then $\nabla^\mu ~t_{\mu \nu} ~= ~0$ is  
satisfied by the following solution:
\beq
R_{\mu \nu} (\omega (e)) ~ =&& 2~ \left[\nabla_\mu \nabla_\nu \phi
~- ~ \nabla_\mu \phi ~\nabla_\nu \phi  \right] \cr 
&&+~  g_{\mu \nu}~ \left[\nabla^2 \phi ~  + ~  2~ (\nabla \phi)^2 
~-~ {\frac 1 2}~ \Lambda e^{2\p} \right]\cr 
&&+ ~{\frac 1 8} ~ e^{-4\phi} \left[ H_{\mu \alpha \beta} ~H_{\nu}^{~~\alpha
\beta}~ -~ {\frac 1 3 }~ g_{\mu \nu}~ H_{\alpha \beta \gamma} ~H^{\alpha
\beta \gamma}\right] \cr
 \nabla^\mu ~[e^{-2\phi}~ H_{\mu \alpha \beta}]~~ =&& 0 \label{greq}
\eeq

To  verify that these indeed provide a solution, substitute 
\bn
 R(\omega(e))~=~  6~[(\nabla \p)^2  ~+~\nabla^2\p ]
~-~{\frac 1 {24}}~ e^{-4\p}~ H_{\a\b\g}~H^{\a\b\g}~ - ~ 2~ \Lambda ~
e^{2\p} 
\en
obtained from the first equation  in to (\ref{t1}) to write

\bn
t_{\mu \nu} &=& 2 ~\left[ \nabla_\mu \phi ~\nabla_\nu \phi - \nabla_\mu
\nabla_\nu \phi \right] 
~ + g_{\mu \nu}~ \left[ (\nabla \phi)^2 + 2\nabla^2 \phi 
- {\frac 1 2}~ \Lambda~ e^{2\p}\right] \cr
&&-~{\frac 1 8}~ e^{-4\phi} ~\left[ H_{\mu \alpha \beta}
~H_\nu^{~~\alpha \beta} - {\frac 1 6}~ g_{\mu \nu} ~
H_{\alpha \beta \gamma} ~H^{\alpha \beta \gamma}\right] \label{t2}
\en
It is useful to notice that
\bn
H^{\m\a\b}~\nabla_\m~H_{\n\a\b} - {\frac 1 6}~\nabla_\n ~(H^{\a\b\g}~H_{\a\b\g})
= - {\frac 1 {18}}~H^{\a\b\g}~\nabla_{[\n}~H_{\a\b\g]} ~=~0
\en
where the last step is implied by the identity $\nabla_{[\n}~H_{\a\b\g]}~\equiv~
0$. Then
\beq
\nabla^\m~t_{\m\n}= &2&\left[ \nabla^2\p~\nabla_\n\p + \nabla_\n(\nabla\p)^2
- \nabla^2\nabla_\n \p + \nabla_\n \nabla^2\p \right] - \partial_\n \p
~e^{2\p} \Lambda \cr
&+& {\frac 1 2}~e^{-4\p}~\nabla^\m \p \left[ H_{\m\a\b}~H_{\n}^{~~\a\b}
- {\frac1 6}~ g_{\m\n}~ H_{\a\b\g}~ H^{\a\b\g} \right] \cr
&-& {\frac 1 8}~ e^{-4\p} ~\nabla^\m H_{\m\a\b} ~ H_\n^{~~\a\b} \label{t3}
\eeq
Next use the identity $\nabla^2\nabla_\n \p - \nabla_\n \nabla^2 \p$ $  
=~ R_{\n\l}(\omega(e))~\nabla^\l \p$~ and first equation of (\ref{greq})
to prove
\bn
2~\left[\nabla^2\p~\nabla_\n \p + \nabla_\n (\nabla \p)^2
- \nabla^2\nabla_\n \p + \nabla_\n \nabla^2 \p \right] ~-~ \partial_\n \p
~e^{2\p}~ \Lambda ~~~~~\cr
~~~~~~~~~~~=~ -~{\frac 1 4}~ e^{-4\phi} ~\nabla^\m \p~\left[ H_{\m \a \b}
~H_\n^{~~\a \b} - {\frac 1 3}~ g_{\m \n} ~
H_{\a \b \g} ~H^{\a\b\g}\right]
\en
This when substituted in (\ref{t3}),  yields
\bn
\nabla^\m~t_{\m\n}~=~ -~{\frac 1 8} ~e^{-2\p}~\nabla^\m(e^{-2\p}~H_{\m\a\b})
~H_\n^{~~\a\b} ~=~0
\en
by the  second equation in  (\ref{greq}).

Thus the solution to self-duality constraint is given in terms of 
equations (\ref{greq}) along with the constraint (\ref{Kconst}) which may be 
rewritten as 
\be
\nabla^{[\a} H^{\b]}_{~~\m\n}  
~= ~2~ \partial^{[\a} \p ~H^{\b]}_{~~\m\n} ~-~2~ \partial_{[\m} \p 
~H_{\n]}^{~~\a\b} ~+ ~ \partial^\s \p ~\d_{[\m}^{[\a} 
~H^{\b]}_{~~\n] \s}  \label{Hconst}
\ee
This constraint is consistent with the second  equation in (\ref{greq}).

Now  notice that (\ref{greq})  are the equations of motion of a dilaton $\p$, 
an axion $B_{\m\n}$ and a cosmological constant $\Lambda$
coupled to gravity in a conformally invariant manner.  An effective 
action with linear $ R$ that yields these as its equations of motion is:
\be
S_{eff} = {\frac 1 2} \int d^4x ~e \left[ 
e^{2\phi} \left( e^{2\p} \Lambda  + R (\omega (e)) + 6 (\partial \phi)^2  \right) 
 - {\frac 1 {24}}~ e^{-2\phi}~ H_{\alpha \beta
\gamma}~ H^{\alpha \beta \gamma} \right] \label{eff}
\ee
This action is conformally invariant.
That is, this action  is unchanged under transformations:
\bn
g_{\mu \nu}(x) ~&\rightarrow& ~{g'}_{\mu \nu}(x) ~= ~\Omega^2 (x) ~
g_{\mu \nu}(x) \cr
\phi(x) ~&\rightarrow& \phi '(x) ~~~~= ~\phi(x)~ - ~ln ~\Omega(x) \cr
B_{\mu \nu} ~&\rightarrow & B_{\mu \nu}' ~~~~~= ~B_{\mu \nu}
\en
That is not surprising, because our starting $SU(2)$ gauge theory action 
(\ref{action}) is classically conformally invariant.

Notice that, while (\ref{greq}) are the equations of motion for effective 
action (\ref{eff}), the constraint (\ref{Hconst}) has to be invoked  
additionally to describe the solution to the self-duality 
constraint $^*F_{\m\n}^i~=~ F_{\m\n}^i$. 

If we restore the  constant $\kappa$  of mass dimensions $-1$
introduced in the (\ref{ansatz}), it shall appear in the 
gravity equations (\ref{greq}) and the effective action (\ref{eff}) 
in a way that $\kappa^2$ can be  interpreted as Newton's constant
of gravity.
Though we started with a gauge theory (\ref{action})
with no dimensionful parameter,  Newton's constant
emerges as a dimensionful modulus of the space of solutions in this theory.

Kinetic energy term for the scalar field $\p$ in the
effective action (\ref{eff}) has the  wrong  sign. It really is not 
a physical field, because it can be rotated away by a Weyl scaling of 
the metric by absorbing it into the conformal factor of new scaled 
conformally invariant metric $g_{\m\n}'(x)$ $=~e^{-\p (x)}~g_{\m\n}(x)$
leading to Poincar$\acute{e}$ gravity from conformal action of a scalar
field.
 
Thus we have demonstrated that solution of the  equations of motion, 
in particular the self-duality equation, of a complex $SU(2)$ gauge 
theory leads to the equations of motion of gravity conformally
coupled to a dilaton and an axion and also a cosmological constant.  
It is worth pointing out that the axion field so obtained 
can also be viewed as propagating torsion. 

\subsection{Anti-self-dual solution}

Next let us analyze  anti-self-dual  solution of the equations of motion:
\be
\cR_{\m\n}^{~~~ab} ~+~ \cR^{ab}_{~~~\m\n} ~=~ -~\Sigma_{\m\n}^{~~~ab} \cR
~+~e_{[\m}^{[a}~\cR_{~~\n]}^{b]} \label{R+R}
\ee
which when contracted with $e_\b^{\n}$ yields:
\be
 \cR ~=~0 ~, ~~~ ~~~~~ \cR_{\m\n} ~ =~ \cR_{\n\m} \label{solantidual1}
\ee
This further, from (\ref{R+R}), implies
\be
{\cR}_{\m\n}^{~~~ab} ~+~ {\cR}_{~~~\m\n}^{ab} ~=~ e_{[\m}^{[a}~ \cR^{b]}_{~~\n]}
\label{solantidual2}
\ee
This equation  fixes 12 out of 21 independent components of
$\cR_{\m\n}^{~~~a b}+ \cR^{a b}_{~~~\m\n}$~, rest 9 are undetermined.
On the other hand six components of $\cR_{\m\n}^{~~~a b}
- \cR^{a b}_{~~~\m\n}$~ are fixed leaving 9 undetermined.
Thus constraints  (\ref{solantidual2}),
which are equivalent to the anti-self-dual constraint (\ref{R+R}),
fix 18 of the 36 independent components of $\cR_{\m\n}^{~~~a b}$.
Notice that the self-dual constraints (\ref{master}) and (\ref{SD3})
and the anti-self-dual constraints  (\ref{solantidual2})
fix complimentary components of $\cR_{\m\n}^{~~~a b}$.

Next we define a (traceless) Weyl tensor associated with $\cR_{\m\n}^{~~~ab}$
as
\bn
{\cC}_{\m\n}^{~~~ab} ~= ~\cR_{\m\n}^{~~~ab} ~+~ {\frac 1 3}~ \Sigma_{\m\n}^{~~~ab} \cR
~-~ {\frac 1 2}~ e_{[\m}^{[a}~ \cR_{\n]}^{~~b]}
\en
Then (\ref{solantidual1}) and (\ref{solantidual2}) imply
\be
{\cC}_{\m\n}^{~~~ab} ~+~ {\cC}_{~~~\m\n}^{ab} ~=~ \cR_{\m\n}^{~~~ab} ~+ ~ 
\cR^{ab}_{~~~\m\n} ~-~ e_{[\m}^{[a} ~{\cR}_{\n]}^{~~b]} ~ =~0 \label{C+C}
\ee
Writing $\cR_{\m\n}^{~~~ab} ~=~ R_{\m\n}^{~~~a b}(\omega(e)) 
~+~ r_{\m\n}^{~~~ab}$ as in (\ref{R+r}),   this constraint can be rewritten as:

\be 
2~R_{\m\n}^{~~~\a\b} ~-~\d_{[\m}^{[\a}~R_{\n]}^{~~\b]} ~=~ 
-~(r_{\m\n}^{~~~\a\b} ~+~ r^{\a\b}_{~~~\m\n} ) ~+~ \d_{[\m}^{[\a}~ 
r_{\n]}^{~~\b]} \label{2R}
\ee
Now for the trace and completely antisymmetric parts
of $h_{\m\a\b}$ as defined in (\ref{ansatz}), pairwise symmetric part
of the four-tensor $r_{\m\n\a\b}$ is
\bn
r_{\m\n}^{~~~\a\b} ~+~ r^{\a\b}_{~~~\m\n} ~=~ {\frac 1 4}~e^{-4\p} 
~H_{[\m}^{~~\a\l}~ H_{\n ] \l}^{~~~\b} &-& 2~ \d_{[\m}^{[\a} ~\nabla_{\n]} 
~\nabla^{\b]} \p \cr
~+~ 2~ \d_{[\m}^{[\a} ~\nabla_{\n]} \p ~\nabla^{\b]} \p
 &-& 2~\d_{[\m}^\a ~\d _{\n]}^\b ~(\nabla \p)^2
\en
The anti-self-dual constraint from (\ref{solantidual1})
implies
\be
r_{\m\n} ~- ~ r_{\n\m} ~=~ {\frac 1 {\sqrt{2}}}~\nabla_{\s} \left( e^{-2\phi}
H_{\m\n}^ {~~~\s} \right) ~=~0 \label{r-r}
\ee
and hence 
\beq
r_{\m\n} ~=~&& 2~\left[ \nabla_\m \phi\nabla_\n \phi ~-~ \nabla_\m \nabla_\n \phi
\right]  \cr
~&&-~ g_{\m\n} \left[ 2(\nabla\phi)^2 ~+~ \nabla^2 \phi\right] ~-~ {\frac 1 8 }~
e^{-4\phi} H_{\m\a\b} H_\n^{~~\a\b} \label{r}
\eeq
and 
\be
R (\omega(e))~ =~ -~r ~=~ 6 \left[ (\nabla \phi)^2 ~+~ \nabla^2 \phi \right] ~+~ {\frac 1 8}
~e^{-4\phi} H_{\a\b\g} H^{\a\b\g} \label{R1}
\ee
Further (\ref{solantidual2}) or (\ref{2R}) implies the constraint
\beq
\nabla^{[\a} R^{\b]}_{~~ \n}(\omega(e)) ~=~ - \nabla^\m \left(r_{\m\n}^{~~~\a\b}
  + ~r^{\a\b}_{~~~\m\n} -~ \delta_{[\m}^{[\a} r^{\b]}_{~~\n]} 
+~{\frac 1  2}~ \d_{[\m}^{\a} ~\d_{\n]}^{\b}~r \right) 
\eeq
where we have used the identity satisfied by the Riemann tensor: ~$\nabla^\m
R_{\m\n\a\b} $ $= \nabla_{[\a} R_{\b] \n}$~.~
It can easily be checked that a solution of this constraint is given by
\newpage
\beq
H_{\m\a\b}  &=& 0 \cr 
R_{\m\n} (\omega (e)) &=& -~ r_{\m\n} ~+~ p_{\m\n} \cr
&=& 2\left[\nabla_\m \nabla_\n \phi - \nabla_\m \phi 
\nabla_\n \phi  \right] + g_{\m\n} \left[ \nabla^2 \phi + 2 (\nabla \phi)^2
\right] ~+~p_{\m\n}~~~~\label{ASDR} 
\eeq
where $p_{\m\n}$ is symmetric ($p_{\m\n}= p_{\n\m}$) and traceless
($p \equiv p^\m_{~\m} = 0$) and  satisfies the equation
\bn
\nabla^{[\a}~p^{\b]}_{~~\m} ~-~ \partial^{[\a} \p~ p^{\b]}_{~~\m} +~\d^{[\a}_{\m}
~p^{\b]}_{~~\n} ~\partial^\n \p ~=~0 
\en
The Reimann
tensor  for this solution (\ref{ASDR}) is
\bn
R_{\m\n}^{~~~\a\b} &= &-~r_{\m\n}^{~~~\a\b} ~+~{\frac 1 2}~\d^{[\a}_{[\m}~
p^{\b]}_{~\n]} \cr~
&=& \d_{[\m}^{[\a}
\left( \nabla_{\n]} \nabla^{\b]} \p ~-~ \nabla_{\n]}\p \nabla^{\b]}\p \right)
~+~ \d_{\m}^{[\a} \d_\n^{\b]} ~(\partial \p)^2 +~{\frac 1 2}~\d^{[\a}_{[\m}~
p^{\b]}_{~\n]}
\en
Thus this provides  a solution to the anti-self-dual constraint 
of gauge theory. 
It is possible that there are other
more general solutions for the anti-self-dual case. 

So far we have discussed only pure complex $SU(2)$ complex gauge theory.
Other matter can also be included in this formulation. This can be  
achieved by adding terms made of other representations of the complex 
$SU(2)$. For example, 
we can add Lorentz scalar fields in triplet representation $\P^i$ 
or fermions $\lambda^i$. In particular, we may add these extra fields
in a supersymmetric manner. This would then lead to the equations
of motion of supergravity. We do this in the next section.

%% file: section3
\section{~N~=~1~ supersymmetric complex ~SU(2)~ gauge theory}

Supersymmetric generalization of Einstein gravity in its
simplest form leads to $N = 1$ supergravity.  This theory,
first discovered about thirty years ago, is
described in terms of, besides a set of auxiliary fields,
physical metric field $g_{\m\n}$ and its superpartner, spin $3/2$
gravitino $\psi_\m$ \cite{SUGRA}. In the spirit of Section 2,
we wish to set up a
locally supersymmetric Yang-Mills theory whose equations of
motion admit $N =1$ supergravity equations as a solution.
General super Yang-Mills action  coupled to tetrad and gravitino,
without kinetic terms for them, and the relevant supersymmetric and
other transformation rules, have also been known for a 
long time \cite{FGvanN, PvanN, FT}. 

We need a  supersymmetric
generalization of the conformally invariant action of complex 
$SU(2)$ gauge  theory of the previous section. For this purpose, we 
introduce a complex  $SU(2)$ triplet vector $N~=~1$ superconformal 
multiplet $(A^i_\mu, $ $  \lambda^i,$ $  {\cD}^i)$ where complex 
${\cD}^i$ is the usual auxiliary field. Notice like complex $A^i_\mu$, 
the fermion is also made of two Majorana triplets:~
$\lambda^i ~= ~\lambda^{(1)i}~ + ~i \lambda^{(2)i}$.~ We have
nine complex off-shell degrees of freedom in $A^i_\m$,
three  in ${\cD}^i$, with total of
$12$ complex off-shell bosonic degrees of freedom which is  same
as the number of off-shell degrees in the fermions
$\l ^i$.  We couple this supermultiplet to off-shell fields of
the  background  conformal supergravity Weyl multiplet 
$(e^a_\m, ~ \psi_\mu, ~ B_\mu)$ 
where the last is an  axial vector field. Here we 
have eight  off-shell real degrees of freedom in  the bosonic 
fields, $e^a_\m$ and   $B_\m$, and an equal number in the gravitino 
field $\psi_\m$.  In terms of these fields Lagrangian density 
$\cL$ for the $N~=~1$ super(conformal) Yang-Mills theory is given by:
\beq
 e^{-1}~ {\cal L} ~=~&& -~{\frac 1 4}~ F^{i \mu \nu} ~F_{i \mu \nu} ~- ~{\frac 1
2}~ {\bar {\lambda}}^i ~\gamma^\m~D_\m ({\hat \omega}) ~\lambda^i ~+ ~{\frac 1 2}~
{\cD}^i ~{\cD}^i\cr 
~&&- ~{\frac 1 4}~ {\bar \psi}_\mu ~\sigma^{\alpha \beta}
~\gamma^\mu ~\lambda^i ~[F^i_{\alpha \beta} ~+ ~{\hat F}^i_{\alpha \beta}]
\label{SYM}
\eeq
Here Majorana conjugate of the fermions is given by:~ 
 $({\bar \lambda}^i)_A ~= ~(\lambda^i)^B
{\cal C}_{B A}$ and $({\bar{\psi}}_\mu)_A~=~ (\psi_\m)^B 
{\cal C}_{B A}$  
where ${\cal C}$ is the charge conjugation matrix and $(A,~ B)$ are 
four component Dirac spinor indices. Supercovariant spin connection 
${\hat \omega}_\m^{~~a b}$  contains
 $\psi_\m $-torsion, but not $\l^i$-torsion: 
\beq
 {\hat \omega}_\mu^{~~a b} ~&= &~\omega_\mu^{~~ a b}(e) ~+ ~\kappa_\mu^{~~a b} \cr
 \kappa_{\mu a b} ~&=& ~{\frac 1 4}~ ( {\bar \psi}_a ~\gamma_\mu ~\psi_b
~+~{\bar \psi}_\mu ~\gamma_a ~\psi_b ~- ~{\bar \psi}_\mu ~\gamma_b ~\psi_a)
\label{sc} 
\eeq
Covariant derivative acting on the fermion is:
\bn
D_\mu({\hat \omega}) \lambda^i = (\partial_\mu + {\frac 1 2}~
{\hat \omega}_{\mu}^{~~a b} \sigma_{a b} -{\frac {3i} 4} \gamma_5 B_\mu)
\lambda^i -\epsilon^{i j k}~ A^j_\mu \lambda^k 
\en
and supercovariant field strength
\be
 {\hat F}^i_{\mu \nu} ~=~ F^i_{\mu \nu} ~-~ {\frac 1 2}~ 
{\bar \psi}_{[\mu} \gamma_{\nu]} \lambda^i \label{superF} 
\ee

The action (\ref{SYM}), besides having  complex  $SU(2)$ gauge and
general coordinate invariances, is  
invariant under  local supersymmetric transformations:
\bn
\d A_\m^i~&=&~ {\frac 1 2}~ {\bar \e}\g_\m\l^i~, ~~~~~~\d \l^i~=~ 
-~{\frac 1 2}~(\s^{\a\b}{\hat F}_{\a\b}^i ~+~i\g_5 \cD^i)\e~ \cr
\d \cD^i ~&=&~ -~ {\frac i 2}~{\bar \e} \g_5 \g^\m 
\left({\hat D}_\m({\hat \omega})\l^i
~+~ {\frac i 2}~\g_5 \cD^i \psi_\m \right) \cr
\d e_\m^a ~&=&~ {\frac 1 2}~{\bar \e} \g^a\psi_\m~, ~~~~~\d B_\m ~=~ -~ 
i{\bar \e} \g_5 \phi_\m~, ~~~~~ \d \psi_\m ~=~ D_\m({\hat \omega})\e
\en
where 
\bn
D_\m({\hat \omega}) \e &\equiv & \left(\partial_\m + {\frac 1 2}~ \sigma_{ab}
~{\hat \omega}_\m^{ab} - {\frac {3i} 4} ~\g_5 B_\m\right) \e \cr 
{\hat D}_\m  ({\hat \omega}) ~\l^i &\equiv&D_\mu ({\hat \omega})
~\lambda^i + {\frac 1 2}~\sigma^{\alpha \beta} ~{\hat F}^i_{\alpha \beta}
~\psi_\mu \cr
\phi_\m&\equiv& {\frac 1 3}~ \g^\n \left( D_\n ({\hat \omega}) \psi_\m 
 - D_\m({\hat \omega}) \psi_\n + {\frac 1 {2e}}~ \g_5 ~\e_{\m\n\a\b} 
~D^\a ({\hat \omega}) \psi^\b \right) \cr 
D_\m ({\hat \omega}) \psi_\n &\equiv& \left(\partial_\m + {\frac 1 2} ~\sigma_{ab}
~{\hat \omega}_\m^{ab} - {\frac {3i} 4} ~\g_5 B_\m\right) \psi_\n + 
{\G}_{\m\n}^{~~~\l} \psi_\l
\en
The action is also  invariant under conformal transformations:
\bn 
&&{e'}^a_\m~ = ~\Omega~ e^a_\m~,~~ ~~~ {\psi '}_\mu ~= 
~\Omega^{\frac 1 2} ~\psi_\m,
~~~~~ {B'}_\m ~= ~B_\m~,  \cr 
&&{A'}_\m^i~ =~ A^i_\m~,~~ ~~~~ {\l'}~ = ~{\Omega}^{-\frac 3 2}~\l^i~, 
~~~~~~{\cD'}^i ~=~ \Omega^{-2}~{\cD}^i
\en  
There is an additional invariance under so called R-symmetry, a local axial
$U(1)$ (assciated  gauge field is  $B_\m$):
\bn
\d A^i_\m =  \d {\cD}^i = \d e^a_\m = 0~, ~~~ \d \l^i = 
{\frac {3i} 4} \a \g_5 \l ^i~, 
 ~~ \d \psi_\m = {\frac {3i} 4} \a \g _5 \psi_\m~ ,~~ \d B_\m = 
\partial_\m \a
\en
Finally action is  also invariant
under  local superconformal transformations:
\bn
&& \d e_\m^a ~=~ 0~, ~~~~~~~\d \psi_\m ~=~ -\g_\m \eta~, ~~~~~~ \d B_\m ~=~ 
i{\bar \eta} \g_5 \psi_\m  \cr
&& \d A_\m^i ~ =~ 0~, ~~~~~~~\d \l^i ~ =~ 0~, ~~~~~~~~~~~~\d {\cal D}^i ~=~0 
\en

As in Section $2$, we have a complex action.  
There are no kinetic terms for the
tetrad field $e_\m^a$, its super partner Majorana $\psi_\m$ and
the  auxiliary axial gauge field $B_\m$. 

\subsection{Equations of motion}

Variation of the action with respect to various fields
($A_\m^i$,  $\l^i$,  $B_\mu$,  $e_\m^a$ and  $\psi_\mu$) 
leads to the following  equations of motion:

\beq
\delta A^i_\mu:& &{\cal D}_\mu ~(F^{i \mu \nu} + {\bar
\lambda}^i ~\gamma^\alpha ~\sigma^{\mu \nu} ~\psi_\alpha) ~=~{\frac 1
2}~\epsilon^{ijk} ~{\bar \lambda}^j ~\gamma^\nu ~\lambda^k \cr
\delta \lambda^i:& & \sl{\hat D}  ({\hat \omega})
~\lambda^i ~=~ 0 \cr 
\delta B_\mu:& & {\bar \lambda}^i ~\gamma_5 ~\gamma_\mu
~ \lambda^i ~ =~ 0  ~~~~~~~\Rightarrow ~~~ \g_5~\l^i~=~ \pm~\l^i \cr
\delta \psi_\mu:& & \sigma^{\alpha \beta} ~{\hat F}^i_{~\alpha
\beta} ~\gamma^\mu ~\lambda^i ~=~ 0 \cr
\delta e^a_{\mu}:&& T_{\mu \nu} ~\equiv ~[F^i_{\mu \alpha}
+ {\bar \lambda}^i ~\gamma^\beta ~\sigma_{\mu \alpha} ~\psi_\beta]
~F^{i~\alpha}_{~\nu} \cr
&& ~~~~~~~~~~~~-~{\frac1 4}~ g_{\mu \nu} ~[F^{i \alpha
\beta} + {\bar \lambda}^i ~\gamma^\rho ~\sigma^{\alpha \beta} ~\psi_\rho]
~F^i_{\alpha \beta} ~= ~0 \label{sol} 
\eeq
where we have used the earlier equations in simplifying 
last two equations and the derivative ${\cD}_\mu$ in the first
equation is covariant
with respect to both the complex $SU(2)$ gauge transformations and
general coordinate transformations. While variation with respect
to the gauge field $A_\m^i$ and the fermions $\l^i$ yield genuine
equations of motion, those with respect to the
fields $B_\m$, $\psi_\m$ and tetrad $e_\m^a$ give only constraints.

We now try to solve these equations. It is straight forward to
check that  the last three equations in (\ref{sol})  are solved by:
\bn
\gamma_5~ \lambda^i ~=~ \mp~ \lambda^i~,  ~~~~~~~~~~
 F^i_{\mu \nu} ~+ ~{\bar \lambda}^i ~\gamma^\alpha
~\sigma_{\mu \nu} ~\psi_\alpha ~= ~\pm ~^*F^i_{\mu \nu}
\en
These in turn imply, a generalized self-duality or anti-self-duality
constraint equation for the supercovariant field strength of
(\ref{superF}):
\be 
\gamma_5~ \lambda^i ~= ~\mp ~\lambda^i~, ~~~~~~
~{\hat F}^i_{\mu \nu} = \pm ^*{\hat F}^i_{\mu \nu} \label{SDSUSY}
\ee 
These constraints make the $\delta A^i_{\mu}$  equation of motion
(the first equation in (\ref{sol}) above) hold
identically. As in the non-supersymmetric case of Section 2, 
for configurations satisfying these constraints, the $N=1$ super 
Yang-Mills Lagrangian density (\ref{SYM}) is a total divergence:
$\cL ~=$ $\mp~ (e/4) F^i_{\m\n} {~}^* {F}^{i\m\n}$ $ =~\mp~ \partial_\m J^\m$.

There is a supersymmetric generalization of Urbantke type formulae
(\ref{U}) as:
\be
 g^{-\frac 1 4 } ~ g_{\mu \nu} ~=~ ( det~ {\hat \phi}_{i j})^{-\frac 1 2}
~{\hat X}_{\mu \nu}, ~~~~~
g^{\frac 1 4} ~g^{\mu \nu} ~=~ ( det~ {\hat \phi}_{i j})^{-\frac 1 2}
~{\hat Y}^{\mu \nu} \label{SU}
\ee
\noindent where quantities ${\hat \p}_{ij}$, ${\hat X}_{\m\n}$ and 
${\hat Y}^{\m\n}$ are given in terms of
the self-dual  or anti-self-dual supercovariant field strength as
\bn
{\hat \phi}_{i j} ~= ~\pm~{\frac 1 2} ~\epsilon^{\mu \nu \alpha \beta} ~
{\hat F}^i_{\mu \nu}~ {\hat F}^j_{\alpha \beta}
\en
\noindent and
\bn
{\hat X}_{\mu \nu} &=& \pm~{\frac 2 3}~ \epsilon^{\alpha \beta \sigma \delta}
~\epsilon^{ i j k} ~ {\hat F}^i_{\mu \alpha} ~{\hat F}^j_{\beta \sigma}~ 
{\hat F}^k_{\delta \nu} \cr
 ~{\hat Y}^{\mu \nu} &=& {\frac 1 3}~ \epsilon^{\mu \alpha \beta \gamma}
~\epsilon^{\lambda \nu \rho \sigma} ~\epsilon^{i j  k} ~{\hat F}^i_{\beta \gamma}
~{\hat F}^j_{\alpha \lambda}~ {\hat F}^k_{\rho \sigma}
\en

To develop solutions of the constraint equations, we next write 
\be
{\hat F}^i_{\mu \nu} ~ =~  
{\hat {\cal R}}_{\mu \nu}^{~~~\alpha \beta} ~\Sigma^i_{\alpha \beta} \label{R}
\ee
We trade three complex fermions $\l^i$ with six Majorana fermions
$\l^{\a\b}$ through the relation $\lambda^i~ =~ \lambda^{\alpha \beta}~\Sigma^i_{\alpha \beta} $. Then
the supercovariant $ {\hat {\cal R}}$-tensor introduced 
above can be written as:
\be
{\hat \cR}_{\m \n}^{~~~\a\b}~=~ {\cal R}_{\mu \nu}^{~~~\alpha \beta} -
{\frac 1 2} ~
{\bar \psi}_{[\mu} \gamma_{\nu]}
\lambda^{\alpha \beta} \label{superR}
\ee
where $ \cR_{\m \n}^{~~~\a\b}$ is the same tensor as introduced in Section 2:
~~ $F^i_{\m\n}~=~ $ $\cR_{\m\n}^{~~~\a\b}~\Sigma_{\a\b}^i$.

Self-duality  or anti-self-duality of the supercovariant field strength (\ref{SDSUSY}) implies: 
\be
 ^*{\hat {\cal R}}_{\mu \nu}^{~~~\alpha \beta} ~=~
\pm~{\tilde {\hat {\cal R}}}_{\mu \nu} ~^{\alpha \beta} \label{SDSUSY1}
\ee
where as earlier $^*$ represents duality with respect to the first pair of
indices $[\m\n]$ and tilde ~$\tilde{}$~ is duality with respect to the
second pair $[\a\b]$. As in Section 2, this equation in turn  leads to
the supersymmetric  generalization of  equation (\ref{SD2}): 
\be
\pm~{\hat \cR}_{\mu \nu}^{~~~a b} ~ -~ {\hat\cR}^{a b}_{~~~\mu \nu} ~ = ~
\Sigma^{~~~a b}_{\mu \nu} ~{\hat \cR}~  - ~ e^{[a}_{[\mu} ~
{\hat\cR}^{b]}_{~~\nu]} 
\ee 
 
We shall consider here only the  self-dual case. In this case
the constraint equation  above is equivalent to the following two 
independent equations
\beq
{\hat {\cal R}}_{\mu \nu} ~+~{\hat {\cal R}}_{\nu \mu}~~ =& 
~{\frac 1 2}~ g_{\mu \nu} ~{\hat \cR} ~~~~~~~~~~~~~~~~~~~~~~~~~~~~~~&(a)\cr 
{\hat\cR}_{\mu \nu}^{~~~a b} ~ -~ {\hat \cR}^{a b}_{~~~\mu \nu} ~ =& ~
{\frac 1 2}~e_{[\m}^{[a}~ \left( {\hat \cR}_{\n]}^{~~b]} ~ -~
{\hat \cR}^{b]}_{~~\n]} \right) ~~~~~~~~~~~&(b)\label{masterSUSY}
\eeq
These fix 18 of the 36 independent components of ${\hat \cR}_{\m\n}^{~~~a b}$.

\subsection{A self-dual solution of equations of motion: ~N~=~1 supergravity}

To solve the self-duality constraint and also the fermion equation of
motion above,  we parametrize:
\be
A^i_\mu ~= ~ {\hat \omega}_\m^{~~\a\b} ~\Sigma^i_{\a\b}~\equiv~ 
[\omega_\mu^{~~\alpha \beta}(e) ~+~\kappa_\mu^{~~\alpha
\beta}]~ \Sigma^i_{\alpha \beta} \label{N1AFIELD}
\ee
with contortion tensor ~$\kappa_{\mu \alpha \beta}$ as given in equation
(\ref{sc}).  Then 
\bn
{\cR}_{\m\n}^{~~~\a\b}~ \equiv~  R_{\m\n}^{~~~\a\b}({\hat 
\omega}) ~=~ R_{\m\n}^{~~~\a\b}(\omega (e))~+~s_{\m\n}^{~~~\a\b}
\en
where
\bn
s_{\m\n}^{~~~\a\b}~=~ \nabla_{[\m} ~\k_{\n]}^{~~~\a\b}~+~\k_{[\m}^{~~~\a\l}~
\k_{\n]\l}^{~~~~\b} 
\en
For $s_{\m\n}~=~g^{\a\b}~s_{\m\a\n\b}$ and $s~=~ g^{\m\n}~s_{\m\n}$ we have
\bn
s_{\m\n}~&=&~ \nabla_\m \k_\n ~-~ (\nabla_\a ~-~ \k_\a)~\k_{\m\n}^{~~~\a} ~-~ 
\k_{\m\a\b}~\k^{\a\b}_{~~~\n} \cr
s~&=&~ 2~\nabla \cdot \k ~-~\k^\m~\k_\m ~-~ \k_{\m\a\b}~\k^{\a\b\m}
\en
where $\k_\m~=~ \k_{\a\m}^{~~~\a}$. Straight forward calculation yields:
\beq
{\hat \cR}_{\m\n}^{~~~\a\b}  - {\hat \cR}^{\a\b}_{~~~\m\n} ~= && {\frac 1 2} 
~{\bar \psi}^{[\a} ~\g^{\b]}~ (\psi_{\m\n} + \l_{\m\n} ) ~-~ {\frac 1 2}
~{\bar \psi}_{[\m} ~\g_{\n]} ~(\psi^{\a\b}  +  \l^{\a\b} ) \cr
&& ~+ ~{\frac 1 8} ~ g^{\s[\a} ~{\bar \psi}^{\b]} ~\g_{[\s} ~\psi_{\m\n]} ~-~ 
{\frac 1 8}~
g_{\s[\m} ~{\bar \psi}_{\n]} ~\g^{[\s} ~\psi^{\a\b]} ~~~~~~~ \label{sg(R-R)}
\eeq
where $\psi_{\mu \nu} $~$ =~ D_{[\mu}({\hat \omega})~ 
\psi_{\nu]}$
$~\equiv~  \partial_{[\m}\psi_{\n]}$ $ + ~ {\frac 1 2}~\sigma_{ab}~
{\hat \omega}_{[\m}^{~~ab} ~\psi_{\n]}$ $ -~{\frac {3i} 4}~\g_5 B_{[\m} 
~\psi_{\n]}~$.  From this we have
\bn
{\hat \cR}_{\m\n} ~ -~ {\hat \cR}_{\n\m} ~ =~ -~ {\frac 1 4}~ {\bar \psi}_{[\m}
~\g^\s ~\left( \psi_{\n\s]} ~ +~ \l_{\n\s]} \right) ~+~{\frac 1 4}~{\bar \psi}^\s
~\g_{[\m} ~\l_{\n\s]} \label{R-R} 
\en

Next like in the last section, we use the  generalized  master equation 
(\ref{masterSUSY}a) 
to construct an expression for $~
[ R_{\mu \nu} (\omega(e))~$ $  - ~{\frac 1 2} ~ g_{\mu \nu} ~R(\omega (e))]$
$~=~-~t_{\mu \nu}$. 
This can easily be seen to be:
\beq
t_{\mu\nu} ~=  &&{\frac 1 2}~ \left[ s_{\m\n} + s_{\n\m} - {\frac 1 2}~
 g_{\m\n} ~s \right]  + {\frac 1 4}~g_{\m\n} ~R(\omega(e)) ~~~~~~\cr 
&&-~ {\frac 1 4}~\left[ {\bar \psi}_{[\m}~\g_{\a]}~\l_\n^{~~\a}
+ {\bar \psi}_{[\n}~\g_{\a]}~\l_\m^{~~\a} 
-{\frac 1 2}~ g_{\m\n}~{\bar \psi}_{[\a}~\g_{\b]}~\l^{\a\b} \right] ~~~~ 
\eeq
We  seek solutions of ~$\nabla^\mu ~t_{\mu \nu}~ = ~0 $.  
Along with this the fermion equation
${\hat {\sl D}}({\hat \omega}) \lambda^i =0$ with constraint
$\gamma_5 \lambda^i = -\lambda^i$ is also to be solved. 
Finally the solution is given by $\l_{\m\n}~= ~-\psi_{\m\n}$ and the
following  set of equations 
\beq
B_\mu = 0~, ~~~~~ \gamma_5 ~\g^\n ~^*\psi_{\a \n} ~=~ 0~, 
~~~~~ R_{\mu \nu}({\hat \omega}) ~ = ~{\frac 1 2}~ {\bar \psi}^\alpha 
~\gamma_5 ~\gamma_\mu ~^*\psi_{\nu \alpha} ~~~  \label{SGEQ}
\eeq
Notice the second equation implies  $\l_{\m\n} ~=$ $ -~\psi_{\m\n}$
$ =~ \g_5~ ^*\psi_{\m\n}$ and also $\g_{[\m}~\psi_{\n\a]} $ $=~0$.
These in turn make right hand side of (\ref{sg(R-R)}) identically zero: 
\beq
{\hat \cR}_{\m\n}^{~~~\a\b} ~= R_{\m\n}^{~~~\a\b} ~+~ {\frac 1 2} 
\psi_{[\m} \g_{\n]} \psi^{\a\b} ~= ~{\hat \cR}^{\a\b}_{~~~\m\n} ~=~
{~}^* {\hat {\tilde {\cR}}}_{\m\n}^{~~~\a\b}
\label{SGR}
\eeq  
The last equation in (\ref{SGEQ}) can be rewritten as
\bn
{\hat \cR}_{\m\n} ~\equiv~ R_{\m\n} ({\hat \omega}) ~-~ {\frac 1 2}~ 
{\bar \psi}_{[\m}
~\g_{\a]}~\l_{\n}^{~~\a} ~ =~ 0 
\en
Both the constraints (\ref{masterSUSY}) are satisfied.
Thus (\ref{SGEQ}) then provide a solution to the generalized
self-duality constraint. Also the fermion equation of motion of 
super Yang-Mills theory is satisfied.
To verify that  is so, using the equations (\ref{SGEQ}), (\ref{SGR})
above and the implied equation ~ 
 $\g^\a \sigma_{\l\r} ~{\hat \cR}_{\a\m}^{~~~\l\r} $$~=~0$,
it can be checked that gravitino field strength $\psi_{a b} 
\equiv e_a^\m e_b^\n \psi_{\m\n}$ satisfies
the following equation: 
\beq
 \slD({\hat \omega})\psi_{ab} &&
~ - ~~~~{\frac 1 2}~\g^\m \sigma^{cd}~ {\hat \cR}_{cd a b}~\psi_\m  \cr
&& =~~\g_c \left(\k_{[a}^{~~~cd}  +  \k^{cd}_{~~~[a} \right)
\psi_{b]d}
 -~ {\frac 1 4}~\g^e\sigma_{cd}~ \psi_{[a}{\bar \psi}_b \g_{e]}~\psi^{cd}~
~=~ 0 ~~~~~~~~~~~~
\eeq
where second step follows by Fierz rearrangement.
This equation is equivalent to the $\l^i$   equation of motion in (\ref{sol}).

Equations (\ref{SGEQ}) can  readily be recognized as the equations 
of motion of $N=1$ Poincar$\acute{e}$ supergravity \cite{SUGRA, PvanN}. These  
describe the dynamics of  Poincar$\acute{e}$ supermultiplet of  physical fields 
$e^a_\m$ and $\psi_\m$ and  auxiliary axial vector field $B_\m$ 
and are governed by an effective linear $R$ Lagrangian density:
\beq
e^{-1} ~{\cal L}_{eff} =~ {\frac 1 2} ~R(e,{\hat \omega}) - {\frac
1 {2e}}~\epsilon^{\mu \nu \alpha \beta} ~{\bar \psi}_\mu ~\gamma_5
~\gamma_\nu ~D_\alpha ({\hat \omega}) ~\psi_\beta 
+ {\frac 3 4}~B^\mu ~B_\mu
\eeq 

Thus starting from the $N~=~1$ supersymmetric complex $SU(2)$ gauge theory
we have obtained a solution of its equations of motion which is
described by $N~=~1$ Poincar$\acute{e}$ supergravity equations of motion. 
Clearly this is not the most general solution of the 
self-dual case. A more general solution would include supermultiplets of
dilaton and axion coupled to the gravity supermultiplet along with
cosmological constant in a (super)conformally invariant manner 
as a complete  supersymmetric
generalization of the gravity solution of the self-duality constraint
obtained in Section 2. Such
a solution, though more involved,  can be developed by the same 
method as described above. It would exhibit  all the symmetries,
including conformal and superconformal symmetries, of the starting
action (\ref{SYM}).

%% file: section4
\section{~N~=~2~ supersymmetric complex ~SU(2)~ gauge theory}

Next level of supersymmetric generalization of Einstein gravity is $N=2$
supergravity \cite{N2SG1, N2SG2, PvanN, FT}. As earlier, this is to 
be obtained from
the self-dual sector of conformally invariant  $N=2$ supersymmetric 
complex $SU(2)$ Yang-Mills theory.  The $N=2$ Yang-Mills
multiplet consists of  two complex $SU(2)$ triplet fermion fields, each 
made up of two Majorana fermions, $\Psi^{iI} = \Psi^{(1)iI} +i \Psi^{(2)iI}$ 
($I= 1,~2$), containing eight off-shell complex triplet fermionic degrees 
of freedom.  We shall split their left and right handed chiral components:
$\l^i_I$ $= (1/2) ~(1+\g_5) \Psi^{iI}$ and $\l^{iI}$ $ = (1/2) ~(1- \g_5) \Psi^{iI}$
so that $\g_5 \l^i_I $ $ = \l^i_I$ and $\g_5 \l^{iI} $ $= - \l^{iI}$.
There is additional compact $SU(2)$ symmetry (which along with a  $U(1)$ 
axial symmetry forms the R-symmetry group) which acts on the upper and
lower chiral indicies $I$ so that chirality and transformation properties
under this real $SU(2)$ are in direct correspondence.  
An equal number of off-shell bosonic degrees of
freedom consist of (1) a complex gauge 
field $A_\mu^i$, (2) two scalar fields, $X^i= X^{(1)i} + i X^{(2)i}$ and its 
charge conjugate ${\bar X}^i = X^{(1)i} - i X^{(2)i}$ (both $ X^{(1)i} $ 
and $X^{(2)i}$ are complex) and (3) symmetric auxilliary field $Y^i_{IJ} $ $ = 
Y^{(1)i}_{IJ} + $ $ i Y^{(1)i}_{IJ} = Y^i_{JI} $ and its conjugate
${\bar Y}^{i IJ}  $ $ = Y^{(1)iIJ} -iY^{(2)iIJ} $ $  = 
\epsilon^{IK} \epsilon^{JL} (Y^{(1)i}_{KL}$ $  - i Y^{(2)i}_{KL})$.
We need a conformally invariant supersymmetric action 
coupling these fields to the background $N=2$ off-shell 
superconformal gravity multiplet.  This background supermultiplet 
contains $24$ off-shell fermionic degrees of freedom
consisting of two Majorana gravitinos  with chiral components
$\psi^I_\m$ and $\psi_{I\m}$ ( $\g_5 \psi^I_\m = \psi^I_\m$, ~  
$\g_5 \psi_{I\m}$ $= - \psi_{I\m}$~) and  additional  Majorana fermion fields 
 with  chiral components $\p^I$ and $\p_I$ ( $\g_5 \p_I $ $= \p_I$, $\g_5 \p^I$
$=- \p^I$~). Equal number of bosonic degrees of freedom are contained in
the tetrad $e^a_\m$, antisymmetric  $T_{\m\n}^{-IJ}$ 
(anti-self dual in $\m, \n$ and antisymmetric in $I, J$) and 
its charge conjugate self-dual $T^{+\m\n}_{IJ}$, a scalar field
$f$, and  an antihermetian  gauge field $V^{~I}_{\m~~J}$$ = 
(V_{\m I}^{~~J})^*$
$ = -V ^{~~~I}_{\m J}$ 
~~($V^{~I}_{\m ~~I} $ $=0$) and an axial gauge field 
$B_\m$ of the associated real $SU(2)$ and $U(1)$ of R-symmetry group.

Action for general $N=2$ super Yang-Mills theory in $N=2$ superconformal
gravity background has been worked out in reference \cite{N2SYM, FT}.
Complex  $SU(2)$  super Yang-Mills Lagrangian density $\cL $ is given by:
\beq
e^{-1} {\cal L} ~=&& -~D_\m{\bar X}^i ~D^{\m} X^i ~ + ~ 
2~f {\bar X}^i X^i ~ +~
{\frac  1 8}~Y^i_{IJ}~ {\bar Y}^{i IJ} ~+~ (\e^{ijk}~ X^j {\bar X}^k)^2 \cr 
&&-~ {\frac 1 4}~ ({\hat F}^{i+}_{\m\n})^2 
+~ X^i {\hat F}^{i\m\n} T^{+}_{\m \n IJ} 
\e^{IJ}  ~-~ {\frac 1 2}~ X^i X^i (T^+_{\m\n IJ} \e^{IJ})^2 \cr
&&+~ 2~{\bar \p}_I \l^{iI} X^i  
~-~ {\frac 1 2}~ {\bar \l}^{iI}  {\sl D} \l^i_I  
-~ \e^{IJ} ~ \e^{ijk}~ {\bar \l}^i_I {\bar X}^j \l^k_J \cr 
&&-~2~ {\bar \l}^{iI} \g _\m \psi_\n^J T^{+ \m\n}_{~~~IJ} X^i
~+~ {\bar \psi}^I_\m {\sl  D} {\bar X}^i \g^\m \l^i_I \cr 
&&-~ \e^{ijk}~ {\bar \l}^{iI} \g^\m \psi^J_\m ~\e^{~}_{IJ} ~X^j 
~{\bar X}^k ~+~ {\bar \psi}^\m_I \psi^\n_J
T^{+~IJ}_{\m\n} X^i {\bar X}^i \cr
&&+~ {\frac 1 2}~\left({\bar \l}^{iI} \g_\m \psi^J_\n~ 
\e^{~}_{IJ} ~ +~ {\bar \psi}_{I\m} \psi_{J\n} ~\e^{IJ} 
~X^i \right) ~^*{\hat F}^{i\m\n}  \cr
&&-~{\frac 1 {2e}} ~\e^{\m\n\r\s}~{\bar \psi}_{I\m} \g_\n \psi^I_\r D_\s X^i 
{\bar X}^i  ~-~ {\frac 1 4} {\bar \l}^{iI} \g^\m \g^\n \psi_{I\m} 
{\bar \psi}^J_\n \l^i_J \cr
&&-~{\frac 1 4} ~\e^{IK} ~\e^{JL}~ \left(
{\bar \psi}_{I\m} \s^{\m\n} \psi_{J\n}  
{\bar \l}^i_K \l^i_L ~-~  {\bar \psi}_{I\m} \psi_{J\n}
{\bar \l}^i_K \s^{\m\n} \l^i_L \right) \cr
&&-~{\frac 1 {2e}}~ \e^{\m\n\r\s}~ {\bar \psi}_{I\m} \g_\n \psi^J_\r
\left( {\bar \psi}^I_\s \l^i_J ~-~ \d^I_J {\bar \psi}^K_\s \l^i_K \right) 
{\bar X}^i \cr
&&+~ {\frac 1 {8e}}~ \e ^{\m\n\r \s}~ {\bar \psi}_{I\m} \psi_{J\n}
~\e^{IJ} \e^{KL}~ \left( 2 {\bar \psi}_{K\r} \g_\s \l^i_L  ~ +~ 
{\bar \psi}_{K\r} \psi_{L\s} X^i\right)X^i  \cr
&&+~  c.~c. \label{N2YMACTION}
\eeq
where the supercovariant  complex $SU(2)$ gauge field strength is
\beq
{\hat F}^i_{\m\n} ~=~ F^i_{\m \n} ~ -~ \left( {\frac 1 2}~{\bar \psi}_{I[\m} 
\g_{\n]} \l^i_J ~\e^{IJ} ~+~ {\bar \psi}_{I\m} \psi_{J\n} 
~\e^{IJ} ~ X^i ~ +~ c.~c. \right) \label{N2SFS}
\eeq
and $F^{i\pm}_{\m\n} \equiv 1/2 ~(F^i_{\m\n} ~\pm~ ^*F^i_{\m\n})$ are 
self- and anti-self-dual combinations of the field strength.
The covariant derivatives  of  scalar fields are:
\beq
&&D_\m X^i ~=~ \left(\partial_\m ~-~ {\frac i 2}~B_\m\right)X^i
~-~ \e^{ijk}~ A^j_\m X^k \cr
&&D_\m {\bar X}^i ~=~ \left(\partial_\m ~+~ {\frac i 2}~B_\m\right)
{\bar X}^i ~-~ \e^{ijk}~ A^j_\m {\bar X}^k \label{CDX} 
\eeq
and those for fermions are
\beq
&&D_\m \l^i_I ~=~ \left(\partial_\m  + {\frac 1 2}~\s^{ab} 
{\hat \omega}_{\m ab} - {\frac i 4}~ B_\m\right)\l^i_I - \e^{ijk}~
A^j_\m \l^k_I  + V^{~~J}_{\m I} \l^i_J  ~~~~~~~~\cr
&&D_\m \l^{iI} ~=~ \left(\partial_\m  + {\frac 1 2}~\s^{ab}
{\hat \omega}_{\m ab} + {\frac i 4}~ B_\m\right) \l^{iI}- 
\e^{ijk}~ A^j_\m \l^{kI}  + V^{~I}_{\m~J} \l^{iJ} ~~~~~~\label{CDL}
\eeq
The supercovariant spin connection  contains the $\psi_\m^I$ torsion and is
\beq
&&{\hat \omega}_{\m ab} ~=~ \omega_{\m ab} (e) ~ + \kappa_{\m ab} \cr
&&\kappa_{\m ab} ~=~ {\frac 1 4}~ \left( {\bar \psi}^I_\m 
\g_a \psi_{Ib}
~-~ {\bar \psi}^I_\m \g_b\psi_{Ia} ~ +~ {\bar \psi}^I_a \g_\m 
\psi_{Ib} ~ +~ c.~c. \right) \label{N2COVSPIN}
\eeq
Here $c.~c.$ stands for charge conjugation which for various fields
acts as: $X^i \leftrightarrow {\bar X}^i$, $Y^i_{IJ} \leftrightarrow 
{\bar Y}^{iIJ}$,
${\hat F}^{i+}_{\m\n} \leftrightarrow {\hat F}^{i-}_{\m\n}$,
$T^{+}_{\m\n IJ} \leftrightarrow T^{-~IJ}_{\m\n}$, $\l^{iI} \leftrightarrow
\l^i_I$, $V_{\m~J}^{~I} \leftrightarrow (V_{\m ~J}^{~I})^* ~=~
V_{\m I}^{~~J}$, $\psi_{I\m}
\leftrightarrow \psi^I_\m$, $\p^I \leftrightarrow \p_I$ and also
$e\rightarrow e^* =-~e$.
Further, it is useful to introduce a generalized supercovariant 
complex $SU(2)$ gauge field strength:
\beq
{\cal F}^i_{\m\n} ~=~  {\hat F}^i_{\m\n} ~ -~ X^i~ T^+_{\m\n IJ} ~
\e^{IJ} ~-~ {\bar X}^i~T^{-~IJ}_{\m\n} ~\e_{IJ} \label{N2GFS}
\eeq

Like in earlier sections, we introduce the tensors ${\cR}_{\m\n}^{~~\a\b}$~, 
${\hat {\cR}}_{\m\n}^{~~~\a\b}$ and in addition the fermionic tensor fields
$\l^I_{\m\n}$~, $\l_{I\m\n}$ and bosonic  tensor fields  $\p_{\m\n}$~,
${\bar \p}_{\m\n}$ through
\beq
&&F^i_{\m\n} ~= ~ {\cR}_{\m\n}^{~~~\a\b}~\S^i_{\a\b}~,~~~~~~~~ {\cF}^i_{\m\n} ~
= ~{\hat {\cR}}_{\m\n}^{~~~\a\b}~\S^i_{\a\b}~,  \cr
&&\l^i_I~~=~~ \l_I^{\m\n} ~\S^i_{\m\n}~, ~~~~~~~~~~~~\l^{iI} ~=~ \l^{I\m\n}~
\S^i_{\m\n}~, \cr
&&X^i~~=~~ \p^{\m\n}~ \S^i_{\m\n}~, ~~~~~~~~~~~~{\bar X}^i ~=~ {\bar \p}^{\m\n}~
\S^i_{\m\n}
\eeq 
Then from the generalized supercovariant field strength (\ref{N2GFS}),
we have the generalized covariant curvature tensor as
\beq
{\hat \cR}_{\m\n}^{~~~\a\b} &=& \cR_{\m\n}^{~~~\a\b} ~~~~~~~~~~~~~\cr
&&~ - \left[ {\frac 1 2}
{\bar \si}^I_{[\m} \g_{\n]} \l^{J\a\b} \e_{IJ}  +
 {\bar \si}^I_\m \si^J_\n \e_{IJ}~\p^{\a\b}    
+T^{~-IJ}_{\m\n}\e_{IJ} {\bar \p}^{\a\b}  + c.~c. ~\right] \cr
&&                   \label{N2GR}
\eeq
where $(\p_{\m\n})^{c.c.} ~=~ {\bar \p}_{\m\n}$ and $ (\l_{I\m\n})^{c.c.} ~=
~ \l^I_{\m\n}$.

\subsection{Equations of motion}

Variations of the action with respect to  fields $f$, $\p_I$,
$\p^I$,  $T^+_{\m\n IJ}$ and $T^{-IJ}_{\m\n}$ yield the following
equations respectively:
\bn
&&~~~X^i{\bar X}^i ~ = ~0~,  ~~~~~~~~~~\l^i_I~{\bar X}^i ~=~0~, ~
~~~~~~~~~~~ \l^{iI}~ X^i ~=~ 0 ~, ~~~~~~~~~~~~~~~~~~~~\cr
&X^i &\left[ {\hat F}^{i+}_{\m\n}  - X^i
T^+_{\m\n IJ}~\e^{IJ}
+ {\frac 1 2} \left( {\bar \psi}_{I\m} \psi_{J\n} \e^{IJ}\right)^+
{\bar X}^i
-{\frac 1 2} \left({\bar \l}^{iI} \g_{[\m} \psi^J_{\n]} \e_{IJ}
\right)^+ \right] 
=~0 ~~~~~~~~~~\cr
&{\bar X}^i &\left[ {\hat F}^{i-}_{\m\n}   
- {\bar X}^i T^{-IJ}_{\m\n}\e_{IJ}
~+ {\frac 1 2}~ \left({\bar \psi}^I_\m \psi^J_\n \e_{IJ}
\right)^- X^i
- {\frac 1 2}~\left( {\bar \l}^{i}_I \g_{[\m} \psi_{\n]J} \e^{IJ}
\right)^- \right]
 =~0 ~~~~~~~~
\en 
These equations have two sets of solutions:
\beq
(i) ~~~~~ \l^i_I ~=~0~, ~~~~~ X^i~=~0~, ~~~~~{\hat F}^{i-}_{\m\n}
~=~ {\bar X}^i~ T^{-~IJ}_{\m\n} ~\e^{~}_{IJ} \label{N2SDSOL} 
\eeq
\beq
(ii) ~~~~ \l^{iI} ~=~0~, ~~~~~ {\bar X}^i ~=~ 0~, ~~~~~
{\hat F}^{i+}_{\m\n} ~=~ X^i~ T^+_{\m\n IJ} ~\e^{IJ} \label{N2ASDSOL}
\eeq
The generalized supercovariant complex 
$SU(2)$ gauge field strength ${\cal F}^i_{\m\n}$ introduced in 
(\ref{N2GFS}) is self-dual and anti-self-dual respectively 
for these two solutions.

Variations of the action with respect to  fields $B_\m$,
$V_{\m~J}^{~I}$,   gauge field $A^i_\m$,  tetrad $e^a_\m$ and
gravitinos $\si^I_\m$, $\si_{I\m}$  are identically
zero when solution (\ref{N2SDSOL}) or (\ref{N2ASDSOL})
is used. In particular, the stress-energy tensor $T_{\m\n}$
obtained by variation with respect to $e^a_\m$ is zero for
the two solutions (\ref{N2SDSOL}) and (\ref{N2ASDSOL}). 

For the case of self-dual solution (\ref{N2SDSOL}), variation
of the action with respect to $\l^{iI}$ and ${\bar X}^i$ are
also identically zero. But variations with respect to
$\l^i_I$ and $ X^i$ yield additional equations of motion for
the fermion field $\l^{iI}$ and the scalar field ${\bar X}^i$.
The fermion equation of motion is:
\beq
\sl{{\hat D}} \l^{iI} ~-~2{\bar X}^i \p^I ~ =~ 0 \label{N2FEQ} 
\eeq
where supercovariant derivative  of the fermion is
\beq
{\hat D}_\m \l^{iI} ~\equiv~ D_\m \l^{iI} ~ -~ {\frac 1 2}~
\s^{ab} {\hat F}^{i+}_{ab} \psi_{J\m} \e^{IJ} ~-~  \sl{\hat D}
{\bar X}^i \psi^I_\m 
\eeq
and   supercovariant derivative of the scalar field is
\beq
{\hat D}_\m {\bar X}^i ~\equiv~ D_\m {\bar X}^i ~-~ {\frac 1 2}~
{\bar \l}^{iJ} \psi_{J\m}
\eeq
The scalar field equation of motion is:
\beq
&&D_a {\hat D}^a {\bar X}^i ~-~ {\frac 1 2}~{\bar \si}_{Ia}{\hat D}^a \l^{iI}
~+~ {\frac 1 2}~\e^{ijk} {\bar \si}^I_a \g^a \l^{jJ} \e^{~}_{IJ} {\bar X}^k 
~ -~ {\frac 1 2}~ {\bar \l}^{iI} \g_a \si^J_b T^{+ ab}_{~~IJ} ~~\cr
&&+~ {\frac 1 2} ~\e^{ijk} {\bar \l}^{jI} \l^{kJ} \e^{~}_{IJ} ~ +~ {\frac 1 2}~
{\hat F}^i_{ab} T^{+ab}_{~~IJ} \e^{IJ} ~+~ {\frac 1 2} ~{\bar \l}^{iI}
~(\p^{~}_I ~+~ \s^{a b} \si_{Iab}) ~~\cr
&&+~\left[2f+ {\bar \si}^a_I \g_a \p^I - {\frac 1 4}~ {\bar \si}^I_a \g_b~
^*\si^{ab}_I - {\frac 1 4} ~{\bar \si}_{Ia} \g_b~ ^*\si^{Iab}\right]{\bar X}^i
~=~0 ~\label{N2SEQ}
\eeq
where  derivative with the Lorentz index  $D_a ~=~ e^\m_a D_\m$
and the supercovariant gravitino field strengths are:
\beq
&&\si_{I\m\n} ~\equiv~ {\hat D}^{~}_{[\m} \si_{\n]I} ~\equiv~ D_{[\m} \si_{\n]I}
~-~ \g^\s T^+_{IJ\s[\m} \si^J_{\n]} \cr
&&\si^I_{\m\n} ~~\equiv~ ~{\hat D}_{[\m} \si^I_{\n]} ~\equiv~~ D_{[\m} \si^I_{\n]}
~-~ \g^\s T^{-IJ}_{~\s[\m} \si^{~}_{\n]J}
\eeq

On the other hand, for the anti-self-dual case (\ref{N2ASDSOL}), 
variation of the  action with respect to $\l^i_I$ and $ X^i$
is identically zero and those with respect to $\l^{iI}$ and ${\bar X}^i$ yield
equations of motion for the fermion $\l^i_I$ and the scalar field
$ X^i$ which are the conjugate versions of the equations
(\ref{N2FEQ}) and (\ref{N2SEQ}) above:
%\newpage 
\beq
&&\sl{\hat D} \l^i_I ~-~ 2 X^i \p_I ~=~0 \cr
&& ~~~ \cr 
&&D_a {\hat D}^a  X^i ~-~ {\frac 1 2}~{\bar \si}^I_a{\hat D}^a \l^i_I
~+~ {\frac 1 2}~\e^{ijk} {\bar \si}_{Ia} \g^a \l^j_J \e^{IJ} X^k ~ 
-~ {\frac 1 2}~ \l^i_I \g^a \si^b_J T^{-IJ}_{ab} ~~~~~~\cr
&&+~ {\frac 1 2} ~\e^{ijk} {\bar \l}^j_I \l^k_J \e^{IJ} ~ +~ {\frac 1 2}~
{\hat F}^{iab} T^{-IJ}_{ab} \e_{IJ} ~+~ {\frac 1 2} ~{\bar \l}^i_I
~(\p^I ~+~ \s^{a b} \si^I_{ab}) ~~~~~~\cr
&&+~\left[2f+ {\bar \si}_a^I \g^a \p_I - {\frac 1 4}~ {\bar \si}^I_a \g_b~
^*\si^{ab}_I - {\frac 1 4} ~{\bar \si}_{Ia} \g_b~ ^*\si^{Iab}\right] X^i
~=~0 ~~ ~~
\eeq
where the supercovariant derivatives are
\bn
&&{\hat  D}_\m \l^i_I ~\equiv~ D_\m \l^i_I ~-~ {\frac 1 2}~ \s^{ab} 
{\hat F}^{i-}_{ab} \si^J_\m \e_{IJ} 
~-~ \sl{\hat D} X^i \si_{I\m} \cr
&& {\hat D}_\m X^i ~\equiv~ D_\m X^i ~ -~ {\frac 1 2}~{\bar \l}^i_I  \si^I_\m
\en

As in earlier sections, for configurations obeying  self-dual 
(\ref{N2SDSOL}) or anti-self-dual
(\ref{N2ASDSOL}) solutions, the Lagrangian density 
(\ref{N2YMACTION}) is a total divergence:
$\cL~=~ \pm (e/4)~ ^*F^{i\m\n}F^i_{\m\n} ~=~ \pm~ \partial_\m J^\m$.

\subsection {N~=~2~ supergravity as a self-dual solution of 
the equations of motion}

We wish to solve the self-duality constraints (\ref{N2SDSOL}) and associated
fermion equation (\ref{N2FEQ}) and  scalar field equation (\ref{N2SEQ}).
We parametrize the complex gauge field $A^i_\m$ as in (\ref{N1AFIELD}): 
$ A^i_\mu ~= ~ {\hat \omega}_\m^{~~\a\b} ~\Sigma^i_{\a\b}$ where now 
${\hat \omega}_\m^{~~\a\b}$ is the $N~=~2$ supercovariantized spin connection  
given by (\ref{N2COVSPIN}). Self-duality implies the same two 
independent constraints (\ref{masterSUSY}) as in the $N~=~1$ case but with
the tensor ${\hat \cR}_{\m\n}^{~~~\a\b}$ now given by the generalized
$N=2$ expression (\ref{N2GR}). Notice for this solution  $\l^I_{\m\n}$ and
${\bar \p}_{\m\n}$ are  self-dual and $\l_{I\m\n}$ and $\p_{\m\n}$ are
anti-self-dual. 

An analogous calculation to that for equation (\ref{sg(R-R)})
of the $N=1$ theory here yields the relation:
\beq
{\hat \cR}_{\m\n}^{~~~\a\b} &-& {\hat \cR}^{\a\b}_{~~~\m\n} ~~~  \cr
&= &{\frac 1 2}~
\left[ {\bar \si}^{[\a}_I \g^{\b]}_{~} (\si^I_{\m\n} +\e^{IJ} \l_{J\m\n})
~-~ {\bar \si}_{I[\m} \g_{\n]} (\si^{I\a\b} + \e^{IJ} \l^{\a\b}_J) 
\right] ~~~~~\cr
&&~~+~ {\frac 1 8}~ \left[ g^{\s[\a}_{~}{\bar \si}_I^{\b]} \g^{~}_{[\s} \si^I_{\m\n]}
~-~ g_{\s[\m} {\bar \si}_{\n]I} \g^{[\s} \si^{\a\b]I}  \right] \cr
&&~~+~ \left[ {\bar \si}_{I\m} \si_{J\n} (T^{-\a\b IJ} - \e^{IJ} \p^{\a\b})
-{\bar \si}^\a_I\si^\b_J (T^{-IJ}_{~\m\n} - \e^{IJ} \p_{\m\n}) 
\right] \cr
&&~~-~\left[T^+_{\m\n IJ} \e^{IJ} \p^{\a\b} ~-~ T^{+\a\b}_{~~IJ} \e^{IJ} \p_{\m\n}
\right] 
~~+~~c.~c. \label{N2R-R}
\eeq 
$N=2$ generalization of the equation (\ref{R-R}) of  $N=1$ case is: 
\beq
{\hat \cR}_{\m\n} ~-~ {\hat \cR}_{\n\m} ~= &-&{\frac 1 4} \left[ 
{\bar \si}^{~}_{I[\m}
\g^\s \left(\si^I_{\n\s]} + \e^{IJ} \l_{\n\s]J}\right) -
{\bar \si}^{I\s}\g^{~}_{[\m} \l^I_{\n\s]}
\e_{IJ} \right]  \cr
&+& \left[ {\bar \si}^{~}_{I[\m} \si^\s_J \left(T^{-IJ}_{~~\n]\s} - 
\e^{IJ} \p^{~}_{\n]\s}\right) 
~-~ T^{-~\s IJ}_{[\m} {\bar \p}^{~}_{\n]\s} \e_{IJ}\right]  \cr
&+&~~ c.~c. \label{N2R-R'}
\eeq

Self-duality constraints (\ref{N2SDSOL}) are solved  if following
hold: 
\bn
&&B_\m~=~0~, ~~~~
V_{\m~~J}^{~I}~=~0~, ~~~~\p^I ~=~0~, ~~~~\p_I~=~0~, ~~~~f~=~0 \cr
&&\l^I_{\m\n}~=~ \l^{I+}_{\m\n} =~\e^{IJ}~\si_{J\m\n}~,~~~~ \l_{I\m\n}
=~\l^-_{I\m\n} =~ \e_{IJ} ~\si^J_{\m\n} \cr
&& T^{-IJ}_{\m\n} ~= ~\e^{IJ}~ \p_{\m\n}~ =~ {\frac 1 {\sqrt2}}~
{\hat F}^-_{\m\n}~, ~~~T^+_{\m\n IJ}~=~ \e_{IJ} ~{\bar \p}_{\m\n}~=~ {\frac 1
{\sqrt 2}}~{\hat F}^+_{\m\n}
\en
where supercovariant field strength for Abelian gauge field $A_\m$ is
\bn
{\hat F}_{\m\n} ~=~  F_{\m\n}~-~ {\frac 1 {\sqrt 2}}~\left(
{\bar \si}^I_\m \si^J_\n \e_{IJ} ~+~ {\bar \si}_{I\m} \si_{J\n} \e^{IJ} 
\right)~, ~~~~~ F_{\m\n} ~=~ \partial_{[\m} A_{\n]}
\en
Then self-duality equations are satisfied if 
\beq
&&~\g^\m \si^I_{\m\n} ~=~0~, ~~~~~~~~~~\g^\m \si_{I\m\n}~=~0 ~~~~~~~~~~~~~~~\cr
&&{\hat \cR}_{\m\n}~=~ R_{\m\n}({\hat \omega},e)~-~{\frac 1 2}~\left(
{\bar \si}^\a_I \g_\m \si^I_{\n\a}\right. ~+ ~\left.{\bar \si}^{I\a} \g_\m 
\si_{I\n\a}\right) \cr
&&~~~~~~-~{\frac 1 {\sqrt 2}} ~\left
( {\bar \si}^I_\m \si^J_\a {\hat F}^{+~\a}_{~\n} ~+~ {\bar \si}_{I\m} 
\si_{J\a} \e^{IJ} {\hat F}^{-~\a}_{~\n}\right)  
 - ~ 2~ {\hat F}^+_{\m\a} {\hat F}^{-~\a}_{\n} ~=~0 \cr
&& {\hat D}_a{\hat F}^{+a b} \equiv D_a({\hat \omega}){\hat F}^{+a b} -~ 
{\frac 1 {\sqrt 2}}~ {\bar \si}_{Ia}~ \si^{ab}_J \e^{IJ} ~=~0\cr
&&{\hat D}_a{\hat F}^{-a b} \equiv D_a({\hat \omega}){\hat F}^{-a b} -~
{\frac 1 {\sqrt 2}}~ {\bar \si}^I_{a}~ \si^{Jab} \e_{IJ} ~=~0 \label{N2SGEQ} 
\eeq 
These make the right hand sides of equations (\ref{N2R-R}) and
(\ref{N2R-R'}) identically zero.  Other equations of motion of the
Yang-Mills theory are also satisfied. 
It can be checked that equations (\ref{N2SGEQ})
imply the following equations for the gravitino field strengths:
\beq
{\sl D}({\hat \omega})\si^I_{ab} ~-~{\frac 1 2}~ \g^\m \s_{cd}
{\hat \cR}^{cd}_{~~~ab} \si^I_\m ~+~ {\frac 1 {\sqrt 2}}~ \g^\m
{\sl {\hat D}}{\hat F}^{-}_{ab} \si_{J\m} \e^{IJ} ~=~0 \cr
\cr
{\sl D}({\hat \omega})\si_{Iab} ~-~{\frac 1 2}~ \g^\m \s_{cd} 
{\hat \cR}^{cd}_{~~~ab} \si_{I\m} ~+~ {\frac 1 {\sqrt 2}}~ \g^\m
{\sl {\hat D}}{\hat F}^+_{ab} \si^J_\m \e_{IJ} ~=~0  \label{SGGEQ}
\eeq
where
\bn
{\hat \cR}_{ab}^{~~~cd} = R_{ab}^{~~~cd} ({\hat \omega}, e)
+ \left[ {\frac 1 2}~{\bar \si}^I_{[a} \g_{b]} \si_I^{cd}
- {\frac 1 {\sqrt 2}}~ {\bar \si}^I_a \si^J_b \e_{IJ} {\hat F}^{+cd}
 - {\hat F}^-_{ab} {\hat F}^{+cd} + c.~c. \right]
\en
Contracting (\ref{SGGEQ})  with $\S^i_{ab}$, the left hand side of 
first equation is identically zero ($\si^I_{ab}$ is anti-self-dual)
and the second equation is the fermion equation (\ref{N2FEQ})
for  $\p^I~=~0$. It can also be checked that  the field equation
(\ref{N2SEQ}) for scalar field ${\bar X}^i$ is satisfied by 
the above solution.

The self-dual solution (\ref{N2SGEQ})  are  equations
of motion of $N=2$ supergravity  action \cite{N2SG1, N2SG2}:
\bn
e^{-1} \cL_{eff} &=& {\frac 1 2} ~R({\hat \omega}, e)  - {\frac  1 4} 
~ F_{\m\n}~  F^{\m\n}\cr ~~~~~
 &~&-~{\frac 1 {2e}}~ \e^{\m\n\a\b}~\left[{\bar \si}^I_\m \g_\n D_\a 
({\hat \omega}) \si_{I\b} ~-~ {\bar \si}_{I\m} \g_\n D_\a 
({\hat \omega}) \si_\b^I \right] ~~~~~~~\cr
&~&+ {\frac 1 {2\sqrt2}}~\left[ {\bar \si}^I_\m \si^J_\n \e_{IJ}
(F^{+\m\n} + {\hat F}^{+\m\n})  
+ {\bar \si}_{I\m} \si_{J\n} \e^{IJ}  (F^{-\m\n}
+ {\hat F}^{-\m\n} ) \right] 
\en

Clearly a more general solution of the self-dual constarint (\ref{N2SDSOL})
and  associated fermion equation (\ref{N2FEQ}) and  scalar equation
(\ref{N2SEQ}) would involve $N=2$
supermultiplets of dilaton and axion coupled to gravity supermultiplet
in a (super)conformally invariant manner as an $N=2$ generalization of
the gravity solution of  self-dual constraint of Section 2.

%% file: section5
\section{Concluding remarks}

We have presented a gauge theory formulation of gravity based on
complex $SU(2)$ group. The action functional is quadratic in  field 
strength.  Here both the complex gauge field $A^i_\mu$ and the metric
$g_{\mu \nu}$ are varied. There is no dynamics for the metric to start
with. Varying the action with respect to metric gives a constraint
equation: which is solved by self-dual or anti-self-dual field
strengths.  This then relates  metric  to  gauge field.
Einstein gravity equations of motion follow from the self-dual
constraint.
Though the starting action  has only a complex dimensionless coupling,
dimensionful constants, in particular Newton's gravitational constant, 
appear as parameters in the space of solutions.

This theory  has some similarities with
Ashtekar formulation of gravity. But there are some characteristic
differences: (i) The action in Ashtekar approach is linear in field
strength, whereas it is quadratic here.  (ii) The equations of
motion ultimately obtained here are not pure gravity but gravity 
coupled to a dilaton and an axion in a conformally invariant manner. 
(iii) Sympletic structure is 
distinctly different.  Canonical momentum conjugate to gauge field 
$A^i_I$ is not densitized spatial triad as in Ashtekar theory 
($ \kappa^{-2}~e~\Sigma^{itI}) $, but like in ordinary gauge theories, 
it is given by $\Pi^{i I} = \tau ~e ~F^{i tI}$.  However,  unlike 
other ordinary gauge theories, there is an additional constraint 
given by self-duality or anti-self-duality condition of the 
field strength.  Thus the
sympletic structure is different from other gauge theories also.
In fact, this makes the constrained Poisson bracket (Dirac 
bracket) of two gauge fields $A^i_I (t, {\bf x})$ and 
$A^j_J (t, {\bf y})$ non-zero.

The analysis has been  extended to $N =1$ complex $SU(2)$
super Yang-Mills theory. This results in a generalized 
self-duality/anti-self-duality condition for not  ordinary 
gauge field strength 
but for supercovariantized field strength.  Finally for
the self-dual case a  solution of 
equations of motion is given by the equations of  $ N=1$ 
supergravity theory.

The discussion has also been extended to $N= 2$  complex 
$SU(2)$ super Yang-Mills theories. Results are similar to
those for  the $N=1$ case.  The self 
duality/anti-self-duality holds for a 
generalized field strength which not only contains the 
usual gauge field strength and terms
involving fermions but also other fields of the supersymmetric
Yang-Mills  and gravity multiplets. For the self-dual case,
the analysis leads to $N=2$ 
supergravity equations of motion.

This analysis can also be extended to $N=4$ complex $SU(2)$
supersymmetric gauge theory. Here self-duality of a more
complicated generalized supercovariant $SU(2)$ field
strength leads to the equations of motion of $N=4$
supergravity. 

Detail discussion of $N=4$ supergravity
obtained from the self-duality constraint in $N=4$ complex
$SU(2)$ super Yang-Mills theory and one-loop quantum corrections
in such  a theory  will be presented elsewhere.

%% file: acknowledgements
\noindent{\bf Acknowledgements}

\vspace{0.5cm}

Discussions with G. Date, S. Kalyana Rama  and J. Maharana
are gratefully acknowledged.

\vspace{0.5cm}

%% file: gtg.bbl
\begin{thebibliography}{9}

\bibitem{ukm} R. Utiyama: Phys. Rev. {\bf 101} (1956) 1597; \\
T.W. Kibble: J. Math. Phys. {\bf 2} (1961) 212; \\
S. Madelstam: Ann. Phys. (N.Y.) {\bf 19} (1962) 25.

\bibitem{ash} Amitabha Sen: Phys. Lett. {\bf 119B} (1982) 89; \\
A. Ashtekar: Phys. Rev. Lett. {\bf 57} (1986) 2244;\\ 
A. Ashtekar: Phys. Rev. {\bf D36} (1987) 1587.

\bibitem{pleb} J. Plebanski: J. Math. Phys. {\bf 18} (1977) 2511.

\bibitem{riccardo} R. Capovilla, J. Dell, T. Jacobson and L. Mason: 
Class. Quantum Grav. {\bf 8} (1991) 41.

\bibitem{thooft} G. 't Hooft: Nucl. Phys. {\bf B357} (1991) 211.

\bibitem{DeWitt} B. DeWitt, {\it Dynamical Theory of Groups and
Fields}, Gordon and Breach, New York, (1965).

\bibitem{yang} C.N. Yang: Phys. Rev. Lett. {\bf 33} (1974) 445.

\bibitem{stelle} K. Stelle: Phys. Rev. {\bf D 16} (1977) 953; \\
 K. Stelle: Gen. Rel. Grav. {\bf 9} (1978) 353.

\bibitem{hehlrep} F.W. Hehl, J. D. McCrea, E.W. Mielke and Y.
Ne'eman, Physics Reports {\bf 258} (1995) 1.

\bibitem{Rsquare} E.W. Mielke, Gen. Rel. Grav. {\bf 13} (1981) 175; \\
P. Baekler, F.W. Hehl and H.J. Lenzen, Vacuum solutions with double
duality properties of a quadratic Poincar$\acute{e}$ gauge field theory,
in {\it Proceedings of the Second Marcel Grossmann Meeting on General
Relativity}, edited by R. Ruffini, North Holland Publishing Company, 1982;\\
V. Szczyrba, Phys. Rev. {\bf D36} (1987) 351.
 
\bibitem{U} H. Urbantke: J. Math. Phys. {\bf 25} (1984) 2321.

\bibitem{thooft1} G. 't Hooft: Phys. Rev. {\bf D14} (1976) 3432.

\bibitem{SUGRA} D.Z. Freedman, P. van Nieuwenhuizen and S. Ferrara:
Phys. Rev. {\bf D13} (1976) 3214; \\
S. Deser and B. Zumino: Phys. Lett. {\bf 62B} (1976) 335; \\
D.Z. Freedman and P. van Nieuwenhuizen: Phys. Rev. {\bf D14} (1976) 912.

\bibitem{FGvanN} M. Kaku, P.K. Townsend and P. van Nieuwenhuizen: Phys. Rev.
{\bf D17} (1978) 3179; \\
S. Ferrara, M.T. Grisaru and P. van Nieuwenhuizen: 
Nucl. Phys. {\bf B138} (1978) 430; \\
S. Ferrara and P. van Nieuwenhuizen: Phys. Letts. {\bf 78B} (1978) 573.

\bibitem{PvanN} P. van Nieuwenhuizen: Physics Reports {\bf 68} (1981) 189. 

\bibitem{FT} E.S. Fradkin and A.A. Tseyltin: Physics Reports {\bf 119} (1985) 233.


%sectioniv

\bibitem{N2SG1} S. Ferrara and P. van Nieuwenhuizen: Phys. Rev. Letts.
{\bf 37} (1976) 1669.

\bibitem{N2SG2} A.S. Fradkin and M.A. Vasiliev: Nuovo Cimento Letts. {\bf 25}
(1979) 79; \\
B. de Wit and J.W. van Holten: Nucl Phys. {\bf B 155} (1979) 530; \\
A.S. Fradkin and M.A. Vasiliev: Phys. Letts. {\bf 85B} (1979) 47; \\
P. Breitenlohner and M.F. Sohnius: Nucl. Phys. {\bf 165} (1980) 483.

\bibitem{N2SYM} M. de Roo, J.W. van Holten, B. de Wit and A. Van Proeyen: 
Nucl. Phys. {\bf 173} (1980) 175; \\
B. de Wit, P.G. Lauwers, R. Philippe and A. Van Proeyen: Phys. Letts.
{\bf 135B} (1984) 295.

\end{thebibliography}
